\documentclass[aps,pra,nobalancelastpage,superscriptaddress,twocolumn,10pt]{revtex4-2}

\usepackage{color,ifthen,amsthm,amsmath,amsxtra,amsfonts,dsfont,graphicx,bm,tikz,scalerel,wasysym,bbm,graphicx,amsthm, braket,physics,empheq}
\usepackage[colorlinks=true,linkcolor=blue, citecolor=blue, urlcolor=blue, bookmarks]{hyperref}
\usepackage[inline]{enumitem}

\newcommand{\be}{\begin{equation}}
\newcommand{\ee}{\end{equation}}
\newcommand{\bea}{\begin{eqnarray}}
\newcommand{\eea}{\end{eqnarray}}
\newcommand{\bse}{\begin{subequations}}
\newcommand{\ese}{\end{subequations}}

\theoremstyle{plain}
\newcommand{\1}{\mathbbm{1}}

\theoremstyle{plain}

\theoremstyle{plain}

\usepackage{color,ifthen,amsthm,amsmath,amsxtra,amsfonts,dsfont,graphicx,bm,tikz,scalerel,wasysym, physics, bbm,  graphicx}
\usepackage[colorlinks=true,linkcolor=blue, citecolor=blue, urlcolor=blue, bookmarks]{hyperref}
\usepackage{CircuitTikz}

\usepackage{amsthm}

\begin{document}

\title{Nonequilibrium Full Counting Statistics and Symmetry-Resolved Entanglement
from Space-Time Duality}

\author{Bruno Bertini}
\affiliation{School of Physics and Astronomy, University of Nottingham, Nottingham, NG7 2RD, UK}
\affiliation{Centre for the Mathematics and Theoretical Physics of Quantum Non-Equilibrium Systems,
University of Nottingham, Nottingham, NG7 2RD, UK}

\author{Pasquale Calabrese}
\affiliation{SISSA and INFN Sezione di Trieste, via Bonomea 265, 34136 Trieste, Italy}
\affiliation{International Centre for Theoretical Physics (ICTP), Strada Costiera 11, 34151 Trieste, Italy}

\author{Mario Collura}
\affiliation{SISSA and INFN Sezione di Trieste, via Bonomea 265, 34136 Trieste, Italy}

\author{Katja Klobas}
\affiliation{School of Physics and Astronomy, University of Nottingham, Nottingham, NG7 2RD, UK}
\affiliation{Centre for the Mathematics and Theoretical Physics of Quantum Non-Equilibrium Systems,
University of Nottingham, Nottingham, NG7 2RD, UK}

\author{Colin Rylands}
\affiliation{SISSA and INFN Sezione di Trieste, via Bonomea 265, 34136 Trieste, Italy}

\begin{abstract}
  Due to its probabilistic nature, a measurement process in quantum mechanics produces a distribution of possible outcomes. This distribution --- or its Fourier transform known as full counting statistics (FCS) --- contains much more information than say the mean value of the measured observable and accessing it is sometimes the only way to obtain relevant information about the system. In fact, the FCS is the limit of an even more general family of observables --- the \emph{charged moments} --- that characterise how quantum entanglement is split in different symmetry sectors in the presence of a global symmetry. Here we consider the evolution of the FCS and of the charged moments of a $U(1)$ charge truncated to a finite region after a global quantum quench. For large scales these quantities take a simple large-deviation form, showing two different regimes as functions of time: while for times much larger than the size of the region they approach a stationary value set by the local equilibrium state, for times shorter than region size they show a non-trivial dependence on time. We show that, whenever the initial state is also $U(1)$ symmetric, the leading order in time of FCS and charged moments in the out-of-equilibrium regime can be determined by means of a space-time duality. Namely, it coincides with the stationary value in the system where the roles of time and space are exchanged. We use this observation to find some general properties of FCS and charged moments out-of-equilibrium, and to derive an exact expression for these quantities in interacting integrable models. We test this expression against exact results in the Rule 54 quantum cellular automaton and exact numerics in the XXZ spin-1/2 chain. 
\end{abstract}
\maketitle
\textit{Introduction.---}
The connection between symmetries and conservation laws --- which culminates in the celebrated Noether's theorem and the Ward identities~\cite{weinberg1995quantum1, weinberg1996quantum2, weinberg2000quantum3} --- is arguably the most fundamental aspect of our understanding of the physical world. Loosely stated, this connection implies that for any continuous symmetry of a physical system there is an associated conserved quantity, or charge, that remains invariant during the time evolution.  An immediate consequence of this fact is that  --- even when the system is out-of-equilibrium --- the presence of a symmetry implies that the value of the associated charge is fixed. A conserved charge, however, can still show non-trivial fluctuations when restricted to a subsystem~\cite{klich2009quantum,eisler2013full,eisler2013universality,lovas2017full,najafi2017full,collura2017full,bastianello2018from,calabrese2020full,doyon2022ballistic,oshima2023disordered}. In fact, whenever the system is prepared in an out-of-equilibrium state, these charge fluctuations evolve non-trivially in time even in the presence of translational invariance~\cite{tartaglia2022real,parez2021quasiparticle,parez2021exact}.

Because of the special nature of the conserved charge, one can expect the time-evolution of its fluctuations to give universal information about the system's dynamics. To make this statement more quantitative let us consider a one-dimensional quantum many-body system enjoying a global $U(1)$ symmetry generated by a charge $\hat Q$ that can be split as a direct sum ${\hat Q=\hat Q_A\oplus \hat Q_{\bar A}}$ for any spatial bipartition $A {\bar A}$. We then prepare the system in some low-entangled non-equilibrium initial state $\ket{\Psi_0}$, let it evolve according to its own unitary dynamics, and look at the time evolution of the full-counting statistics (FCS) at time $t$
\begin{equation} \label{eq:FCS}
  Z_{\beta}(A,t)= \expval{e^{i \beta \hat Q_A}}{\Psi_t}= \tr \big(\hat \rho_A(t) e^{i \beta \hat Q_A}\big).
\end{equation}
Here $A$ is a contiguous block and $\hat \rho_A(t)=\tr_{\bar A} \ketbra{\Psi_t}{\Psi_t}$ is the reduced density matrix of the subsystem $A$. This quantity characterises the full probability distribution of $\hat Q_A$ in $\ket{\Psi_t}$. Indeed, considering its derivatives in $\beta=0$ one can reproduce all the moments of the reduced charge.  

Because of the generic phenomenon of local relaxation~\cite{PolkovnikovReview, calabrese2016introduction, VidmarRigol, essler2016quench, doyon2020lecture, bastianello2022introduction, alba2021generalized} we expect the FCS~\eqref{eq:FCS} to show qualitatively different behaviours in the two regimes 
\begin{enumerate*}[label=(\roman*)\!] 
  \item\label{it:stat} $t\gg |A|$ and 
  \item\label{it:noneq}$t\ll |A|$,
\end{enumerate*}
where $|A|$ denotes the size of $A$. Specifically, for $t\gg |A|$ we expect the subsystem $A$ to relax to a stationary state $\hat \rho_{\mathrm{st}, A}$ and, therefore, the FCS to become time-independent at leading order in time
\begin{equation}
  Z_{\beta}(A,t) \simeq {\rm tr}[\hat \rho_{\mathrm{st},A} e^{i \beta \hat Q_A}]\,.
  \label{eq:FCSstat}
\end{equation}
For this reason we refer to~\ref{it:stat} as the equilibrium regime. Conversely, in the regime~\ref{it:noneq} the FCS generically shows a non-trivial time dependence even at leading order in time, and we hence refer to it as the out-of-equilibrium regime. 

\begin{figure*}[th!]
  \centering
  \includegraphics[width=\linewidth]{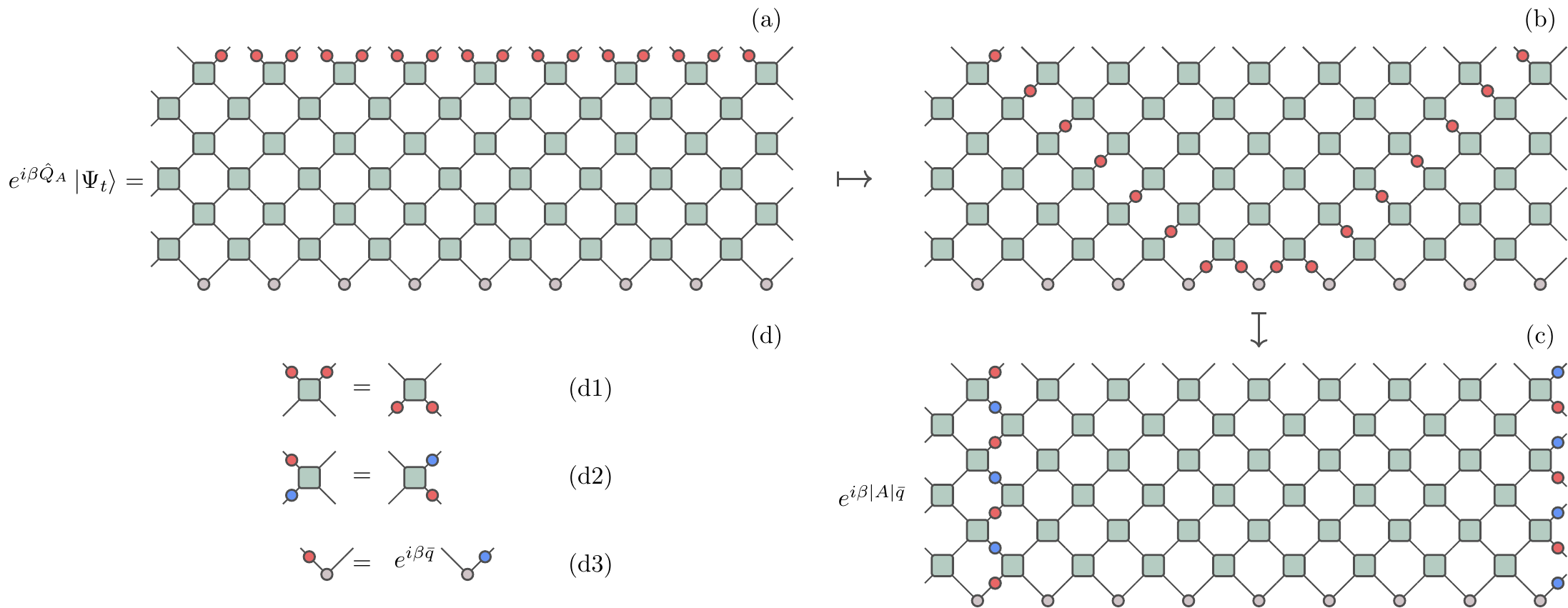}
  \caption{Diagrammatic representation of $e^{i \beta Q_A}\ket{\Psi_t}$. We adopted the convention that when they are acting sideways (cf.\ the diagrams (d2) (d3)) the matrices act from left to right.}
  \label{fig:stringdeformation}
\end{figure*}

In this letter we consider the evolution of~\eqref{eq:FCS} in the out-of-equilibrium regime and obtain two main results. First, we show that, unexpectedly, whenever the state $\ket{\Psi_0}$ is an eigenstate of the charge $\hat Q$ the FCS in the out-of-equilibrium regime can be written in terms of an \emph{equilibrium quantity} for the ``dual system'' where the roles of space and time have been exchanged. This allows us to find a number of general features of its evolution in any locally interacting systems. Second, we use our observation to find an \emph{exact prediction} for the non-equilibrium dynamics of~\eqref{eq:FCS} in interacting integrable models treatable by Thermodynamic Bethe Ansatz (TBA)~\cite{takahashi1999thermodynamics, korepin1997quantum}. To the best of our knowledge, this represents the first closed form expression of the FCS for interacting systems in the out-of-equilibrium regime, and complements existing results on the dynamics of FCS in the local equilibrium state~\cite{myers2020transport,doyon2020fluctuations,doyon2022ballistic,gopalakrishnan2022theory,krajnik2022exact,krajnik2022universal,scopa2022exact}.

In fact, our arguments are not limited to charge fluctuations in a single replica and can be extended to entanglement-related quantities. Namely, they also apply for the more general family of observables known as charged moments (CM)~\cite{goldstein2018symmetry, xavier2018equipartition, parez2021quasiparticle}
\begin{equation}
  Z_{\alpha,\beta}(A,t)= \tr \big( \hat \rho_A(t)^\alpha e^{i \beta \hat Q_A}\big), \qquad \alpha,\beta\in \mathbb R.
  \label{eq:Chargemoments}
\end{equation}
These quantities measure how the entanglement between $A$ and $\bar A$ is decomposed in different charge sectors --- their Fourier transforms in $\beta$ are the symmetry resolved entanglement entropies (SREEs)~\cite{laflorencie2014spin, goldstein2018symmetry, xavier2018equipartition, bonsignori2019symmetry, murciano2020entanglement} --- and, remarkably, they are accessible in ion-trap experiments~\cite{lukin2019probing, azses2020identification, neven2021symmetry, vitale2022symmetry,rath2023entanglement}. 

\textit{Space-time duality.---}To explain our reasoning it is convenient to begin by considering the case in which the system of interest is a \emph{brickwork quantum circuit}. Namely, it is composed of a collection of $2L$ qudits with $d$ internal states arranged on a discrete lattice and its time evolution is implemented by discrete applications of the unitary operator
\be
\hat{\mathbb U} =  {\hat\Pi}^\dag {\hat U}^{\otimes L} \hat \Pi {\hat U}^{\otimes L}.
\ee
Here $\hat U$ acts on two neighbouring sites and $\hat \Pi$ is the periodic shift by one site. Brickwork quantum circuits dispose of most features of real-world quantum matter but retain spatial locality and unitarity. Therefore, they are regarded as the simplest possible extended quantum systems~\cite{nahum2017quantum,chan2018solution,chan2018spectral,fisher2023random}. Importantly, these systems emerge naturally in the context of both classical~\cite{suzuki1991general,schollwock2011density} and quantum~\cite{arute2019quantum} simulation of quantum dynamics.

In a quantum circuit the conservation of the charge $\hat Q$ can be implemented locally via a traceless operator $\hat q$ that together with $\hat{U}$ satisfies
\begin{equation}
  (e^{i \beta \hat q}\otimes e^{i \beta \hat q}) \hat U = \hat U (e^{i \beta \hat q}\otimes e^{i \beta \hat q})\,, \qquad \forall \beta\in\mathbb R\,. 
  \label{eq:chargecons}
\end{equation}
This ensures that $\hat Q=\sum_j \hat q_j$ --- where $\hat q_j$ acts as $\hat q$ at site $j$ and as the identity elsewhere --- is conserved and can be split as a direct sum for any spatial bipartition. 

Analogously, considering the family of two-site translational invariant pair-product states  
\begin{equation}
  \!\!\!\!\ket{\Psi_0}= \ket{\psi_0}^{\otimes L}\!\!\!\!,\,\,\, \ket{\psi_0}=\!\!\sum_{i,j=1}^d m_{ij} \ket{i,j}, \,\,\, {\rm tr}[mm^\dag]=1,
\end{equation}
where $\{\ket{i}\}$ is a basis of the Hilbert space of a single qudit, we have that iff 
\begin{equation}
  e^{i \beta \hat q} \hat m = e^{i \beta \bar q} \hat m e^{- i \beta \hat q^T}\,, \qquad \forall \beta\in\mathbb R\,,
  \label{eq:eigencond}
\end{equation}
with $\bar{q}$ a scalar and $(\cdot)^T$ denoting transposition, then $\hat Q\ket{\Psi_0}=L \bar q \ket{\Psi_0}$.

Introducing the following tensor-network inspired~\cite{cirac2021matrix} diagrammatic representation
\begin{equation}
  \hat U= \begin{tikzpicture}[baseline={([yshift=-0.6ex]current bounding box.center)},scale=0.55]
    \prop{0}{0}{colU}
  \end{tikzpicture},
  \quad 
  \hat m=   \begin{tikzpicture}[baseline={([yshift=-0.6ex]current bounding box.center)},scale=0.55]
    \draw [semithick, rounded corners=1,colLines]  (-0.25,.5) -- (0,0) -- (-0.25,-.5);
    \draw [thick,rounded corners=1,colLines,fill=colIst] (0,0) circle (4.5pt);
  \end{tikzpicture},
  \quad 
  e^{i\beta \hat q} = \begin{tikzpicture}[baseline={([yshift=-0.6ex]current bounding box.center)},scale=0.55]
    \draw [semithick, rounded corners=1,colLines]  (0,.5) -- (0,-.5);
    \draw [thick,rounded corners=1,colLines,fill=myred] (0,0) circle (4.5pt);
  \end{tikzpicture},
  \quad
  e^{-i\beta \hat q^T} = \begin{tikzpicture}[baseline={([yshift=-0.6ex]current bounding box.center)},scale=0.55]
    \draw [semithick, rounded corners=1,colLines]  (0,.5) -- (0,-.5);
    \draw [thick,rounded corners=1,colLines,fill=myblue] (0,0) circle (4.5pt);
  \end{tikzpicture},       
  \label{eq:diagrep}
\end{equation}
we can depict $e^{i \beta \hat Q_A}\ket{\Psi_t}$ as in Fig.~\ref{fig:stringdeformation}(a). Note that the matrices in \eqref{eq:diagrep} act from bottom to top and for convenience we define $|A|$ as the number of qudits in the subsystem divided by two. Our first step is to show that, using Eq.~\eqref{eq:chargecons} and Eq.~\eqref{eq:eigencond}, we can ``deform'' the string of red circles in the diagram passing from Fig.~\ref{fig:stringdeformation}(a) to Fig.~\ref{fig:stringdeformation}(c). 

\begin{figure}[t]
  \centering
  \includegraphics[width=\linewidth]{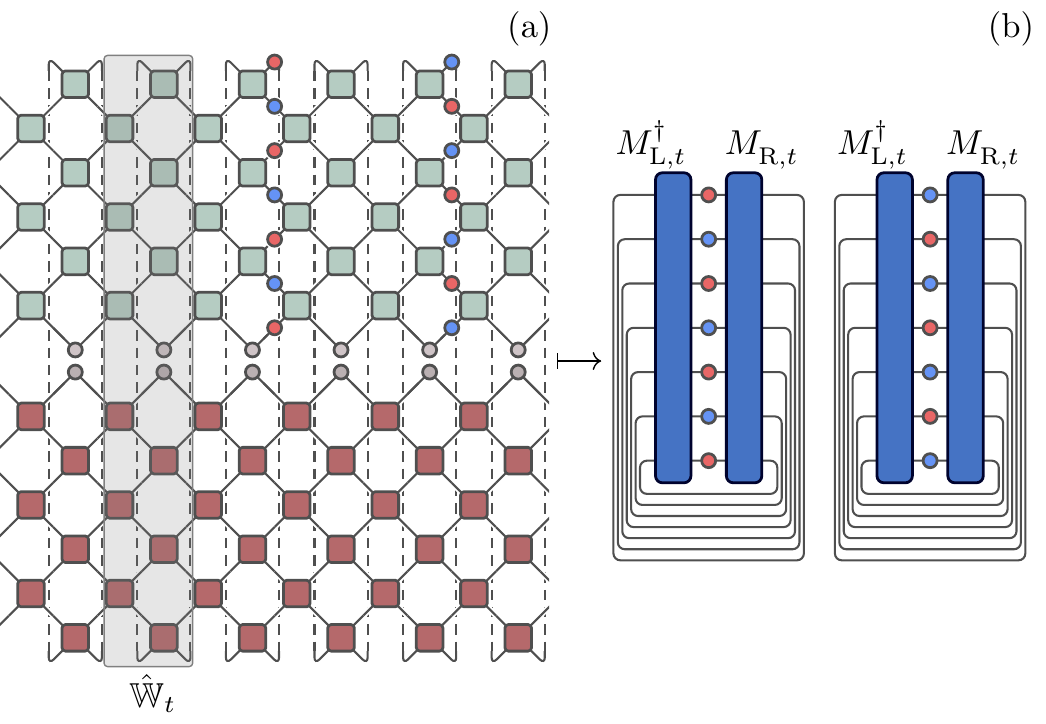}
  \caption{\label{fig:FCS}
  Diagrammatic representation of $e^{-i\beta \abs{A} t} Z_{\beta}(A,t)$ for (a)
  generic choices of $t$, $\abs{A}$, $\abs{\bar{A}}$, and (b) in the regime
  $\abs{A},\abs{\bar{A}}>2t$.  The diagram in the left panel follows directly
  from the definition of time-evolution  (and subsequent
  manipulations in Fig.~\ref{fig:stringdeformation}), but it can be
  equivalently understood as a result of \emph{space propagation} by identifying
  the shaded part as a transfer matrix $\mathbb{W}_t$ that acts on the
  \emph{vertical} lattice of $2t$ qudits (cf.\ Eq.~\eqref{eq:defSpaceEvolution}). 
  Whenever the sizes of the subsystem and the system are large enough compared to the
  time $t$, the action of $\mathbb{W}_t$ in each subsystem can be replaced by fixed
  points $M_{\mathrm{R},t}$ and $M_{\mathrm{L},t}^{\dagger}$, which
  gives the diagram in (b).}
\end{figure} 

To see this we first repeatedly use the diagrammatic representation of Eq.~\eqref{eq:chargecons}, reported in Fig.~\ref{fig:stringdeformation}(d1), and obtain Fig.~\ref{fig:stringdeformation}(b) from Fig.~\ref{fig:stringdeformation}(a). Next, we use the relations~\eqref{eq:chargecons} and~\eqref{eq:eigencond} to propagate the circles ``sideways'', i.e.\ in the space direction. Specifically, Eq.~\eqref{eq:diagrep} implies that $\hat m$ already acts in the space direction while~\eqref{eq:chargecons} gives
\be
\hat{\tilde U} (e^{i \beta \hat q}\otimes e^{-i \beta \hat q^T})= (e^{-i \beta \hat q^T} \otimes e^{i \beta \hat q})\hat{\tilde U}\,, 
\label{eq:chargeconsspace}
\ee
where we introduced the reshuffled local gate with elements $\tilde U^{ij}_{kl}=U^{lj}_{ki}$~\cite{bertini2019exact}. The two relations \eqref{eq:eigencond} and \eqref{eq:chargeconsspace} are represented diagrammatically in Fig.~\ref{fig:stringdeformation}(d2) and Fig.~\ref{fig:stringdeformation}(d3) respectively. In particular, $\hat{\tilde U}$ is still represented by the green tensor in Eq.~\eqref{eq:diagrep} but now the latter is seen as a matrix propagating from left to right. A repeated application of Fig.~\ref{fig:stringdeformation}(d2) and Fig.~\ref{fig:stringdeformation}(d3) brings us from Fig.~\ref{fig:stringdeformation}(b) to Fig.~\ref{fig:stringdeformation}(c).

To conclude, we use the representation in Fig.~\ref{fig:stringdeformation}(c) to compute the FCS via ``space propagation''~\cite{banuls2009matrix,muellerhermes2012tensor,hastings2015connecting, bertini2018exact,bertini2019exact,bertini2022entanglement, bertini2022growth, ippoliti2021postselectionfree, ippoliti2021fractal,lerose2021influence,thoenniss2023efficient}. Namely, we represent Eq.~\eqref{eq:FCS} as in Fig.~\ref{fig:FCS}(a) and contract it from left to right using the transfer matrix $\mathbb{W}_t$ highlighted in the figure. Translating it into an equation we have  
\be \label{eq:defSpaceEvolution}
\!\!Z_{\beta}(A,t) \!=\! e^{i \beta \bar q A} {\rm tr}[{\hat{\mathbb W}_t}^{|\bar A|}  \!(e^{i \beta \hat{\tilde Q}_t}\!\otimes_r\! \1) \mathbb W_t^{|A|} \!(e^{-i \beta \hat{\tilde Q}_t}\!\otimes_r\! \1)],
\ee
where the tensor product $\otimes_r$ is between forward and backward time sheets (top and bottom part of Fig.~\ref{fig:FCS}(a)) and we introduced the charge of the space-time swapped model $\hat{\tilde Q}_t = \sum_{j=1}^{t} (\hat q_{2j-1} - \hat q^T_{2j}))$. Using now that for $x\geq 2t$ the matrix ${\hat{\mathbb W}}^x_t$ becomes a projector onto its unique fixed points parametrised by the matrices $M_{L,t}$ and $M_{R,t}$ (see Fig.~\ref{fig:FCS}(b), and, e.g., Ref.~\cite{bertini2022growth} for more details), we find that for $|A|, |\bar A| \geq 2t$~\footnote[2]{In the quantum circuit, the speed of propagation is $1$ and $t$ is simply the integer number of times the time evolution is appplied.  In the case of the TBA integrable models discussed later with continuous time evolution a (model dependent) velocity scale, $c$ must be introduced and the condition instead reads $A,\bar{A}\gg c t$.},
\begin{equation} \label{eq:dualityFCS}
Z_{\beta}(A,t) = e^{i \beta \bar q |A|} {\rm tr}[\hat{\tilde{ \rho}}_{{\rm st},t} e^{i \beta \hat{\tilde Q}_t}]{\rm tr}[\hat{\tilde \rho}_{{\rm st},t} e^{- i \beta \hat{\tilde Q}_t}],
\end{equation}
where we introduced the pseudo density matrix $\hat{\tilde \rho}_{{\rm st},t}= M_{L,t}^\dag M_{R,t}$~\cite{bertini2022growth}. We now follow Ref.~\cite{bertini2022growth} and interpret $\hat{\tilde \rho}_{{\rm st},t}$ as the stationary state of the ``space-time swapped'' circuit --- i.e.\ the quantum circuit obtained from the starting one by exchanging the roles of space and time. Although this matrix is not Hermitian in the usual sense, i.e., $\hat{\tilde \rho}_{{\rm st},t}\neq \hat{\tilde \rho}^\dag_{{\rm st},t}$, it is diagonalisable. Moreover, its eigenvalues are real, non-negative, and sum to one~\cite{bertini2022growth}. This means that it can be interpreted as a thermal state of a system with a non-hermitian, yet positive, Hamiltonian~\cite{brody2014biorthogonal}.

A comparison between \eqref{eq:FCSstat} and \eqref{eq:dualityFCS} reveals that the FCS in the non-equilibrium regime is written in terms of equilibrium FCS for the space-time swapped model. This means that the FCS in the non-equilibrium regime can be \emph{written in terms of equilibrium quantities}. This observation constitutes our first main result. 

\textit{General Properties.---} Before showing how Eq.~\eqref{eq:dualityFCS} can be used to produce quantitative predictions we make three important observations.
\begin{enumerate*}[label=(\Alph*)]
  \item \label{it:FCS1} The analogue of Eq.~\eqref{eq:dualityFCS} also holds for the CM~\eqref{eq:Chargemoments}.
\end{enumerate*}
Indeed, applying the above reasoning we find
\begin{equation}
\label{eq:chargemomentprediction}
Z_{\alpha,\beta}(A,t) = e^{i \beta \bar q |A|} {\rm tr}[\hat{\tilde \rho}^\alpha_{{\rm st},t} e^{i \beta \hat{\tilde Q}_t}]{\rm tr}[\hat{\tilde \rho}^\alpha_{{\rm st},t} e^{- i \beta \hat{\tilde Q}_t}],
\end{equation}
for $|A|, |\bar A| \geq 2t$.
\begin{enumerate*}[resume*]
\item \label{it:FCS2} Eqs.~\eqref{eq:dualityFCS} and \eqref{eq:chargemomentprediction} immediately imply that SREEs display a \emph{delay-time for activation}, i.e.\ the entanglement entropies of a sector with charge $Q=|A|\bar q+\Delta Q$ is identically zero up to a time $t_{\rm D}\propto|\Delta Q|$.
\end{enumerate*}
This observation generalises the free-fermion result of Ref.~\cite{parez2021quasiparticle, parez2021exact} to generic quantum circuits. To prove it we note that it suffices to show that $\hat \rho_{A,Q}(t)$ --- the density matrix reduced to the block of charge $Q$ --- has zero trace for $t\leq t_{\rm D}$. Indeed, since $\hat \rho_{A,Q}(t)$ is positive semi-definite, it has zero trace precisely when it is zero. Using now Eqs. ~\eqref{eq:FCS} and~\eqref{eq:dualityFCS} and considering the physically relevant case of charge operators with integer spectrum we have  
\begin{equation}
  \mkern-10mu
  \tr[\hat \rho_{A,Q}(t)] \mkern-4mu = \mkern-8mu
  \smashoperator{\int\limits_{-\pi}^{\pi}}
  \mkern-6mu\frac{{\rm d}\beta}{2\pi}
  \!\tr\!\big[\hat{\tilde \rho}_{{\rm st},t} e^{-i \beta  \hat{\tilde Q}_t}\big]
  \!\tr\!\big[\hat{\tilde \rho}_{{\rm st},t} e^{i \beta  \hat{\tilde Q}_t}\big]
  e^{i \beta \Delta Q}.\mkern-8mu 
\end{equation}
Using that the integrand is analytic and $2\pi$-periodic we have that the integration contour can be shifted along the imaginary axis. Therefore, if the integrand vanishes at either $\pm i \infty$ the integral is zero. As is shown in the Supplemental Material (SM)~\cite{Note1}, this happens for $t\leq t_{\rm D}:={|\Delta Q|}/{2 q_\mathrm{diff}}$ where $q_\mathrm{diff}$ is the difference between the largest and smallest eigenvalues of $\hat q$ and is equal to the maximal eigenvalue of $\hat{\tilde Q}_t/t$~\cite{Note1}.  Moreover, using the continuity equation for $
\hat{Q}_A$ it is possible to interpret $\hat{\tilde Q}_t$ as the associated current operator integrated in time, up to $t$  at the left (right) boundary of $A$~\cite{bertini2023dynamics}.  Thus the time delay is the shortest possible time in which the charge $|\Delta Q|$ can be transported through the boundaries of the system.
\begin{enumerate*}[resume*]
\item \label{it:FCS3} Interpreting $\hat{\tilde \rho}_{{\rm st},t}$ as a (generalized) Gibbs state one can use general arguments of statistical mechanics to show that the ``number entropy'' $-\sum_Q \tr[\hat \rho_{A,Q}(t)] \log \tr[\hat \rho_{A,Q}(t)]$ grows in time as $(1/2)\log t$~\cite{Note1}.
\end{enumerate*}
This observation once again generalizes the free-fermion result of Refs.~\cite{parez2021quasiparticle, parez2021exact} to generic systems.

\textit{Integrable models.---}Let us now proceed to show that the general observations above can be used to find  predictions in interacting integrable quantum many-body systems. To this aim, we begin by recalling few basic facts about the latter systems. The spectrum of an integrable model generically consists of a number of stable quasiparticle species,  parameterized by a species index $n$ and a rapidity $\lambda$. Their properties are described through a compact set of kinematic data: energy $\varepsilon_n(\lambda)$, momentum $p_n(\lambda)$, and charge $q_n$ of a quasiparticle, as well as the two-particle scattering kernel $T_{nm}(\lambda)$, and the density of the charge $q_0$ in the reference state without quasiparticles.  In the equilibrium regime,  for a large subsystem  $|A|\to\infty$ we can use the TBA framework along with the  Quench Action method~\cite{caux2013time, caux2016quench}  to find the asymptotic logarithmic density of charged moments
\begin{equation} \label{eq:deq}
  \begin{aligned}
    d_{\alpha,\beta}&=\lim_{\abs{A}\to\infty} 
    \frac{1}{\abs{A}}
    \log \tr[\rho^\alpha_{\mathrm{st},A} e^{i \beta Q_A}]\\
    &=i \beta q_0
    +\sum_{n}\int\!\frac{\mathrm{d}\lambda}{2\pi}
    p^{\prime}_n(\lambda)\mathcal{K}^{(\alpha,\beta)}_n(\lambda),
  \end{aligned}
\end{equation}
with the functions $\mathcal{K}^{(\alpha,\beta)}_n(\lambda)$ satisfying a set
of coupled integral equations
\begin{equation} \label{eq:Keq}
\begin{aligned}
    \mathcal{K}^{(\alpha,\beta)}_n&=
    \mathrm{sgn} [p_n']
    \log\big[(1-\vartheta_n)^\alpha
   +\frac{\vartheta_n^\alpha}{x_n^{\mathrm{sgn} [p_n']}}\big]\\ 
    \mkern-6mu\log[ x_n(\lambda)]&\!=\!
    -i \beta q_n\!+\! \sum_{m}\!\int\mkern-8mu{\rm{d}}\mu\,
    T_{nm}(\lambda-\mu)
  \mathcal{K}^{(\alpha,\beta)}_m(\mu).\mkern-4mu
\end{aligned}
\end{equation}
Here $\vartheta_n(\lambda)$ are the occupation functions of the quasiparticles
in the long time steady state which can be determined exactly for certain
combinations of initial states and
models~\cite{piroli2017what,denardis2014solution,brockmann2014quench,wouters2014quenching,pozsgay2014correlations,bertini2016quantum,bertini2014quantum,piroli2019integrableI,piroli2019integrableI,alba2016the,piroli2016exact,denardis2015relaxation,
mestyan2017exact,rylands2023solution,rylands2022integrable,rylands2019loschmidt}.
This reproduces the results of~\cite{piroli2022thermodynamic} obtained using the Gartner-Ellis theorem.  

To evaluate Eqs.~\eqref{eq:dualityFCS} and \eqref{eq:chargemomentprediction} we need to write the stationary densities of the system where the roles of position and time are swapped. Following Ref.~\cite{bertini2022growth} we obtain them from \eqref{eq:deq} and \eqref{eq:Keq} by performing a space-time swap in Fourier space, i.e.\ exchanging the roles of $p_n(\lambda)$ and $\varepsilon_n(\lambda)$. This leads to
\begin{equation}\label{eq:swappeddensity}
  \begin{aligned}
    s_{\alpha,\beta}&=\lim_{t\to\infty}\frac{1}{t}
  \log \tr\!\big[\tilde\rho^\alpha_{\mathrm{st},t} e^{i \beta \tilde{Q}_t}\big]\\
    &=i\beta\tilde{q}_0 +
    \sum_{n}\int\!\frac{\mathrm{d}\lambda}{2\pi}
    \varepsilon^{\prime}_n(\lambda)\mathcal{L}_n^{(\alpha,\beta)}(\lambda),
  \end{aligned}
\end{equation}
where now we have that
\begin{equation}
\begin{aligned}\label{eq:swappedTBA}
    \mathcal{L}^{(\alpha,\beta)}_n&=
    {\rm{sgn}}[\varepsilon_n']\log\big[(1-\vartheta_n)^\alpha
    +\frac{\vartheta_n^\alpha}{y_n^{{\rm{sgn}}[\varepsilon_n']}}
    \big],\\ 
    \mkern-6mu
    \log[y_n(\lambda)]&\!=\!
    -i\beta\tilde{q}_n\!+\!\sum_m\!\int\mkern-8mu\mathrm{d}\mu\,
    T_{nm}(\lambda-\mu)
    \mathcal{L}_{m}^{(\alpha,\beta)}(\mu).
    \mkern-4mu
\end{aligned}
\end{equation}
The dual driving term $\tilde{q}_n$ and reference-state density $\tilde{q}_0$
depend upon the form of $\hat{\tilde Q}_t$, but can be determined on a case by case
basis.  We now arrive at our second main result: for interacting integrable models the leading order values of CM in the out-of-equilibrium regime are determined by Eqs.~\eqref{eq:swappeddensity} and~\eqref{eq:swappedTBA}. 

We emphasise that Eqs.~\eqref{eq:deq} and~\eqref{eq:swappeddensity} predict an exponential decay in time of the charged moments in the nonequilibrium regime and an exponential decay in space in the equilibrium regime. This behaviour can be understood  intuitively by noting that the logarithm of a charged moment in a stationary state is generically extensive. Interestingly, this phenomenology is in contrast with what observed in the case of random unitary circuits with conservation laws, which show sub-exponential decay~\cite{rakovszky2019sub, huang2020dynamics}. The latter results are not in contradiction with space-time duality: they merely indicate that for random unitary circuits with conservation laws the logarithms of the charged moments in the space-time swapped stationary state are not extensive, i.e.,~$s_{\alpha,\beta}=0$.

\textit{Tests.---}To test this prediction we perform two nontrivial checks, one analytic and one numerical with details on each presented in the supplemental material~\cite{Note1}.  For the analytic check we employ the so-called \emph{Rule 54} quantum cellular automaton~\cite{bobenko1993two} which, despite being an interacting and TBA integrable model~\cite{gombor2022integrable,friedman2019integrable}, is simple enough to allow for the exact calculation of several nonequilibrium quantities~\cite{prosen2016integrability,prosen2017exact,gopalakrishnan2018operator,gopalakrishnan2018hydrodynamics,inoue2018two,friedman2019integrable,alba2019operator,klobas2019time,buca2019exact,alba2021diffusion,klobas2020matrix,klobas2020space,klobas2021exact,klobas2021exactrelaxation,klobas2021entanglement} (see Ref.~\cite{buca2021rule} for a recent review).  Comparing exact results of the charged moments in a quench from a set of solvable initial states we find exact agreement with~(\ref{eq:swappeddensity}, \ref{eq:swappedTBA}).  

\begin{figure}
    \centering
    \includegraphics[width=\linewidth]{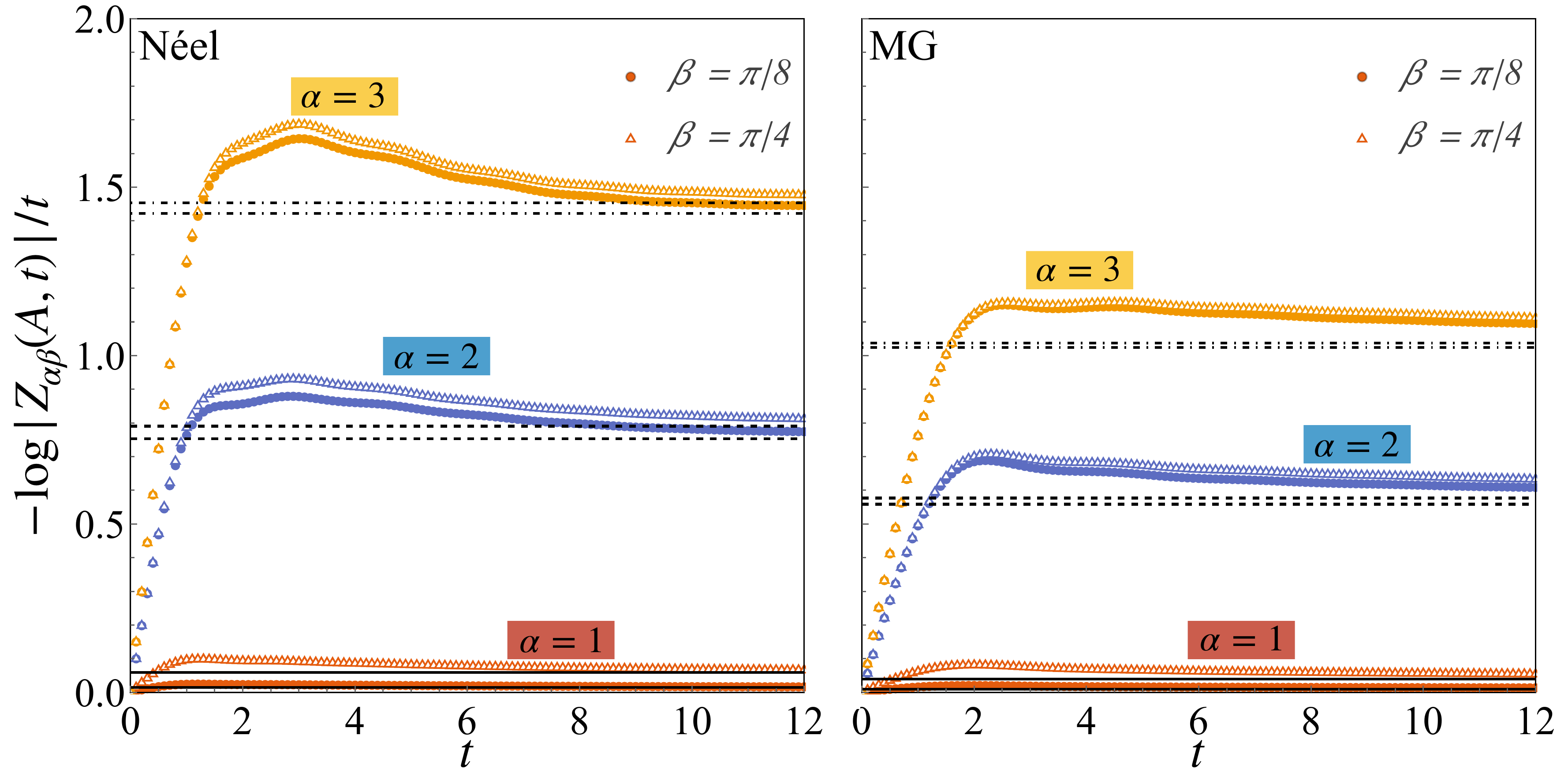}
    \caption{Logarithmic slope of the charged moments after a quench in the XXZ model with $\Delta=1/2$, starting from the N\'eel state (left panel), Majumdar-Gosh state (right panel).  Symbols are the iTEBD data computed with $|A|=50$, straight lines are the asymptotic predictions $i\beta |A| \bar{q}+t(s_{\alpha,\beta}+s_{\alpha,-\beta})$. Different $\alpha$ values correspond to different colours and have been identified with labels; different symbols identify different values of $\beta$.  In both cases spin-flip symmetry fixes these to be real.}
    \label{fig:neel_mg}
\end{figure}

For the numerical check we use the paradigmatic example of an interacting integrable model: the $XXZ$ spin chain,
$
\hat H=\sum_{j=1}^{2L} \hat \sigma^x_{j}\hat\sigma^x_{j+1}+ \hat\sigma^y_{j}\hat\sigma^y_{j+1}+ \Delta \hat\sigma^z_{j}\hat\sigma^z_{j+1},
$
quenched from either the  N\'eel state, $\ket*{\Psi_{\rm{N}}}=\ket{\uparrow\downarrow}^{\otimes L}$, or the Majumdar-Gosh state, $\ket*{\Psi_{\rm{MG}}}=[\left(\ket{\uparrow\downarrow}-\ket{\downarrow\uparrow}\right)/\sqrt{2}]^{\otimes L}$.  We compare Eqs.~\eqref{eq:swappeddensity} and~\eqref{eq:swappedTBA} against numerical simulations using infinite time-evolving block decimation scheme (iTEBD)~\cite{vidal2007itebd} directly in the thermodynamic limit and report the  results in Fig.~\ref{fig:neel_mg} finding good agreement.

\textit{Finite time dynamics.---}In analogy with what happens for R\' enyi entropies~\cite{bertini2022growth}, the expressions \eqref{eq:deq} and \eqref{eq:swappeddensity} show a breakdown of the quasiparticle picture for CM~\cite{parez2021quasiparticle, parez2021exact} in the presence of interactions. More precisely, they imply that a quasiparticle description is only possible if one admits that the quasiparticle velocities depend on both $\alpha$ and $\beta$. This contrasts the usual assumption of the quasiparticles being observable-independent~\cite{calabrese2005evolution,alba2017entanglement}. 
We also remark that, as for the R\'enyi entropies~\cite{bertini2022growth}, combining \eqref{eq:deq} and \eqref{eq:swappeddensity} by assuming abrupt saturation of each mode, one can reconstruct the full dynamics of CM at leading order, i.e.\
\begin{equation} \label{eq:fullconj}
\begin{aligned}
    \log Z_{\alpha,\beta}&(A,t) \simeq 
    i \beta \bar q \abs{A}\\
&+\sum_{n}\int\!{\mathrm{d}\lambda}
    \min(\abs{A},2 t v^{(\alpha,\beta)}_n(\lambda)) d^{(\alpha,\beta)}_n(\lambda),
    \end{aligned}
\end{equation}
where the explicit expression of $v^{(\alpha,\beta)}_n(\lambda)$  and $d^{(\alpha,\beta)}_n(\lambda)$ is reported in the SM~\cite{Note1}. This means that, upon computing the Fourier transform of the CM via saddle point integration, our result gives access to the full dynamics of SREE at leading order. The FCS are an important example of this as the Fourier transform returns the probability of measuring charge $Q$ in $A$ at time $t$. We find that it is normally distributed with standard deviation~\cite{Note1},
\begin{equation}\label{eq:lastEq}
\mathcal{D}(t)=\sum_n\int \!{\rm d}\lambda \,q_{n,{\rm eff}}^2\rho_n(1-\vartheta_n)\text{min}(|A|,2t |v_n|).
\end{equation}
where $v_n$ and $q_{n,{\rm{eff}}}$ are the standard quasiparticle velocity and effective charge.  This shows a remergence of the quasiparticle picture in the $\alpha\to1$ limit as is the case for Renyi entropies~\cite{bertini2022growth}.

\textit{Conclusions. ---} In this Letter we have studied the quench dynamics of full counting statistics and charged moments, which characterise symmetry resolved entanglement, in interacting systems.  Upon identifying two dynamical regimes --- equilibrium and nonequilibrium --- we have shown that both can be analysed using equilibrium techniques via space-time duality. We used this observation to determine some generic features of symmetry resolved entanglement: the presence of a time delay for activation, and the logarithmic growth of the number entropy. Moreover, we have conjectured a closed-form expression for full counting statistics and charged moments in interacting integrable models and tested it against exact analytical and numerical results. We considered global quantum quenches from symmetric initial states, i.e.\ eigenstates of the charge, but our method can be directly applied to mixed initial states relevant for transport settings~\cite{doyon2020fluctuations,myers2020transport,doyon2022ballistic,gopalakrishnan2022theory,krajnik2022exact,krajnik2022universal,scopa2022exact}. An immediate direction for future research is to extend our approach to cases in which the initial state explicitly breaks the $U(1)$ symmetry~\cite{ares2023entanglement} or, more generally, to full counting statistics of non-conserved observables~\cite{groha2018full, collura2018how}.

\begin{acknowledgments}
This work has been supported by the Royal Society through the University Research Fellowship No.\ 201101 (BB), by the Leverhulme Trust through the Early Career Fellowship No.\ ECF-2022-324 (KK), and by the ERC under Consolidator grant number 771536 NEMO (CR and PC).
\end{acknowledgments}

\footnotetext[1]{See the Supplemental Material (SM), which contains Refs.~\cite{huang1987statistical,klobas2023inprep,ilievski2016string,vidal2003mps}. The SM contains: (i) A discussion of the general properties of charged moments in quantum circuits; (ii) An explicit expression of $v^{(\alpha,\beta)}_n(\lambda)$ and $d^{(\alpha,\beta)}_n(\lambda)$ and saddle point calculation of the Fourier transformed charged moments; (iii) A demonstration of equivalence between the exact result and the TBA prediction for Rule 54; (iii) A self consistent summary of the TBA for the gapless XXZ chain; (iv) More details about our numerical experiments.}

\bibliographystyle{apsrev4-2}
\bibliography{bibliography.bib}

\onecolumngrid
\newpage 
\newcounter{equationSM}
\newcounter{figureSM}
\newcounter{tableSM}
\stepcounter{equationSM}
\setcounter{equation}{0}
\setcounter{figure}{0}
\setcounter{table}{0}
\setcounter{section}{0}
\makeatletter
\renewcommand{\theequation}{\textsc{sm}-\arabic{equation}}
\renewcommand{\thefigure}{\textsc{sm}-\arabic{figure}}
\renewcommand{\thetable}{\textsc{sm}-\arabic{table}}

\begin{center}
  {\large{\bf Supplemental Material for\\
  ``Evolution of Full Counting Statistics and Symmetry-Resolved Entanglement from Space-Time Duality''}}
\end{center}
Here we report some useful information complementing the main text. In particular
\begin{itemize}
  \item[-] In Sec.~\ref{sec:details} we discuss some general properties of charged moments in quantum circuits providing an explicit derivation of the observations (B) and (C) that we reported the main text. 
  \item[-] In Sec.~\ref{sec:fullconj} a derivation of Eq.~\eqref{eq:fullconj} of the main text and calculate the Fourier transform of the chareged moments.  
  \item[-] In Sec.~\ref{sec:Rule54} we shown that the prediction agrees with the exact calculation in Rule 54.
  \item[-] In Sec.~\ref{sec:XXZBethe} we provide a self consistent summary of the TBA description of the spin-1/2 XXZ chain for principal root of unity points, i.e., for $\Delta=\cos({\pi}/{(p+1)})$. 
  \item[-] In Sec.~\ref{sec:Numerics} we give details about our numerical experiments in the XXZ chain and provide further comparison plots. 
\end{itemize}

\section{General properties of charged moments in quantum circuits}
\label{sec:details}
In this section we provide a self-contained derivation of the observations (B)and (C) in the main text. Namely, that (B) in any quantum circuit the SREEs display a \emph{delay-time for activation} and (C) the ``number entropy'' grows logarithmically in time. We consider the physically relevant case of charges $\hat Q$ with integer spectrum, e.g., the number operator. 

Let us begin from (B). As discussed in the main text, we can prove it
by showing that the trace of $\hat \rho_{A,Q}(t)$ is zero for
$t\leq t_{\rm D}\propto |\Delta Q|$, with $\Delta Q = Q-\abs{A} \bar{q}$.
Under this assumption we can express the trace of $\hat \rho_{A,Q}(t)$ as
follows (cf.\ Eqs.~\eqref{eq:FCS} and~\eqref{eq:dualityFCS})
\begin{equation}
\tr[\hat \rho_{A,Q}(t)] =
  \int_{-\pi}^{\pi}\!\frac{{\rm d}\beta}{2\pi} 
  \tr\big[\hat{\tilde \rho}_{{\rm st},t} e^{-i \beta  \hat{\tilde Q}_t}\big]
  \tr\big[\hat{\tilde \rho}_{{\rm st},t} e^{i \beta  \hat{\tilde Q}_t}\big]
  e^{i \beta \Delta Q}. 
\label{eq:appintreptr}
\end{equation}
Now we note that, if $\hat Q$ has integer spectrum the same holds for $\hat{\tilde Q}_t$. This means that the function 
\begin{equation}
f(\beta):=
  \tr\big[\hat{\tilde \rho}_{{\rm st},t} e^{-i \beta  \hat{\tilde Q}_t}\big]
  \tr\big[\hat{\tilde \rho}_{{\rm st},t} e^{i \beta  \hat{\tilde Q}_t}\big], 
\end{equation}
is analytic and $2\pi$-periodic in $\beta$. Moreover, since $\Delta Q$ is also integer, the integrand 
\be
g(\beta):=f(\beta) e^{i \beta \Delta Q},
\ee
shares these properties. 

\begin{figure}
\centering
\begin{tikzpicture}[xscale=1, yscale=1]
\draw node[below] at (-5.15,0) {$-\pi$};
\draw node[below] at (-3,-0.05) {$\pi$};
\draw node[below] at (3-0.15,0) {$-\pi$};
\draw node[below] at (5,-.05) {$\pi$};
\filldraw [black] (-5,0) circle (1.6pt);
\filldraw [black] (-3,0) circle (1.6pt);
\filldraw [black] (5,0) circle (1.6pt);
\filldraw [black] (3,0) circle (1.6pt);
\draw[-stealth, line width=0.2mm] (-6,0) -- (-2,0);
\draw[-stealth, line width=0.2mm] (-4,-1.4) -- (-4,1.4);
\draw[-stealth, line width=0.2mm] (2,0) -- (6,0);
\draw[-stealth, line width=0.2mm] (4,-1.4) -- (4,1.4);
\draw[|-latex, line width=0.3mm] (-.35,0.5) -- (.35,0.5);
\draw[->, line width=0.4mm] (-5,0) -- (-3.9,0);
\draw[-, line width=0.4mm] (-4,0) -- (-3,0);
\draw[->, line width=0.4mm] (3,0) -- (3,0.6);
\draw[->, line width=0.4mm] (3,1) -- (4.1,1);
\draw[-, line width=0.4mm] (3,0.5) -- (3,1) -- (5,1);
\draw[->, line width=0.4mm] (5,1) -- (5,0.4);
\draw[-, line width=0.4mm] (5,1) -- (5,0);
\end{tikzpicture}
\caption{Graphical representation of the contour deformation.}
\label{fig:contours}
\end{figure}
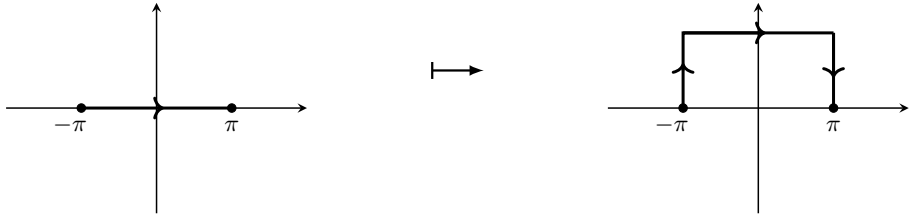

Using the analiticity of $g(\beta)$ we deform the integration contour as
depicted in Fig.~\ref{fig:contours}. Because of the periodicity of the
integrand, the contributions of the two vertical sections cancel each other and
we are left with the horizontal one. This means that, if the integrand vanishes
at either $i \infty$ or $-i\infty$, the integral is zero. To understand when
this happens we consider 
\begin{equation}
  \label{eq:limit}
  \begin{aligned}
    \lim_{z\to i\,\mathrm{sgn}[\Delta Q] \infty }\mkern-10mu
    \frac{\log g(z)}{\abs{z}} 
    =& - |\Delta Q| 
    + \mkern-10mu \lim_{z\to \mathrm{sgn}[\Delta Q]\infty}\mkern-10mu
    \frac{
    \log\left(
    \tr\big[\hat{\tilde \rho}_{{\rm st},t} e^{z  \hat{\tilde Q}_t}\big]
    \tr\big[\hat{\tilde \rho}_{{\rm st},t} e^{-z \hat{\tilde Q}_t}\big]\right)}
    {\abs{z}}
    \leq - |\Delta Q| + 2 t ( q_{\max}- q_{\min})\,.
  \end{aligned}
\end{equation}
Here we used that the maximal eigenvalue of $\hat{\tilde Q}_t$ is $t ( q_{\max}- q_{\min})$ and its
minimal eigenvalue is $-t(q_{\max}- q_{\min})$ where $q_{\max}$ and $q_{\min}$
are the maximal and minimal eigenvalues of $\hat q$. We see that for 
\be
t < \frac{|\Delta Q|}{2 ( q_{\max}- q_{\min}) } =: t_{\rm D}, 
\ee
the limit \eqref{eq:limit} is always negative, implying that the integrand
vanishes exponentially at $i\,\mathrm{sgn}[\Delta Q]\infty$. This
proves~(B). 

To prove~(C) we consider the regime $t\geq t_{\rm D}$ and evaluate
the asymptotic behaviour of the integral~\eqref{eq:appintreptr} for large times
using the saddle point approximation 
\begin{equation}
  \tr[\hat \rho_{A,Q}(t)] \simeq
  \frac{f(i \beta^*)}{\sqrt{2\pi |\log(f)''|_{\beta=i \beta_*}}} e^{-\beta_* \Delta Q},
\label{eq:saddle}
\end{equation}
where $\beta_*$ is the solution of the saddle point equation 
\be
\Delta Q = \expval*{\hat{\tilde Q}_t}_{\beta_*}-\expval*{\hat{\tilde Q}_t}_{-\beta_*},
\label{eq:saddlepointeq}
\ee
and we introduced the shorthand notation $\expval*{O}_\beta$ for the expectation value of the operator $O$ in the state  
\be
\hat \rho_\beta = \frac{\hat{\tilde \rho}_{{\rm st},t} e^{\beta  \hat{\tilde Q}_t}}{{\rm tr}[\hat{\tilde \rho}_{{\rm st},t} e^{\beta  \hat{\tilde Q}_t}\!]}.
\ee
Computing explicitly the derivative we have 
\be
\log(f)''|_{\beta=i \beta_*} = -\expval*{(\hat{\tilde Q}_t-\expval*{\hat{\tilde Q}_t}_{\beta_*})^2}_{\beta_*}-\expval*{(\hat{\tilde Q}_t-\expval*{\hat{\tilde Q}_t}_{-\beta_*})^2}_{-\beta_*} \leq 0. 
\ee

Let us now interpret $\hat \rho_\beta$ as a well defined (generalised) Gibbs state in a system with a large volume $t$ and analyse \eqref{eq:saddle} using standard methods of statistical mechanics~\cite{huang1987statistical}. We begin by rewriting~\eqref{eq:saddle} and~\eqref{eq:saddlepointeq} as follows 
\begin{equation}
\label{eq:statmechz1}
  \tr[\hat \rho_{A,Q}(t)] \simeq 
  \frac{e^{-\Delta Q \beta_*(\Delta Q) }}{\sqrt{2\pi (\sigma_{\tilde Q, \beta_*}^2+\sigma_{\tilde Q,-\beta_*}^2})}
  \left(\frac{Z(\beta_*)}{Z(0)}\right) \left(\frac{Z(-\beta_*)}{Z(0)}\right),
\end{equation}
where $\beta_*=\beta_*(\Delta Q)$ is a function of $\Delta Q$ determined by
\begin{equation}\label{eq:statmechz2}
  \Delta Q = \expval*{\hat{\tilde Q}}_{\beta_*}-\expval*{\hat{\tilde Q}}_{-\beta_*}.
\end{equation}
Here we introduced the partition function
\be
\frac{Z(\beta)}{Z(0)}=\tr\big[\hat{\tilde \rho}_{{\rm st},t} e^{\beta\hat{\tilde Q}_t}\big],
\ee
and denoted by $\sigma_{Q}^2$ the fluctuations of a generic charge $\hat{Q}$. In particular, we expect the fluctuations to scale linearly with the volume~\cite{huang1987statistical}
\be
\sigma_{\tilde Q}^2 \sim t \sigma_{\tilde q}^2. 
\ee 
Considering now small $\Delta Q$ and expanding $\beta_*$
from~\eqref{eq:statmechz1} we find   
\begin{equation}
  \label{eq:expbeta}
  \beta_*(\Delta Q) = \frac{\Delta Q}{2 \sigma_{\tilde Q,0}^2} + O(\Delta Q^2).
\end{equation}
Combining \eqref{eq:expbeta} it with the relations 
\be
\partial_\beta \log Z(\beta) = \expval*{\hat{\tilde Q}}_\beta, \qquad \partial^2_\beta \log Z(\beta) = \sigma_{\tilde Q}^2,
\ee
we finally obtain   
\be
\tr[\hat \rho_{A,Q}(t)] \simeq \frac{1}{\sqrt{4\pi t \sigma_{\tilde q}^2}}\exp[-\frac{\Delta Q^2}{4 t \sigma_{\tilde q}^2}]\,.
\ee
As shown in Ref.~\cite{parez2021exact}, this form of $\tr[\hat \rho_{A,Q}(t)]$
gives the following leading order scaling for the ``number entropy'' 
\begin{equation}
  S_\text{num}=-\sum_Q \tr[\hat \rho_{A,Q}(t)] \log \tr[\hat \rho_{A,Q}(t)] \simeq \frac{1}{2}\log t + O(t^0)\,.
\end{equation}

\section{Full dynamics of charged moments at leading order in time}
\label{sec:fullconj}

By combining the results for the charge moments in both the equilibrium and non-equilibrium regimes along with the idea that each mode experiences an abrupt saturation we can reconstruct the full dynamics of the charge moments to leading order.  To do this we recall some facts about TBA integrable models.  As described in the main text, the spectrum of such models consists of stable quasiparticle excitations labelled by an index $n$ and rapidity $\lambda$.  A stationary state is thus specified by its quasiparticle content and in the thermodyanmic limit is given by a set of functions $\rho_n(\lambda),\rho_n^t(\lambda)$ and $\vartheta_n(\lambda)$ which are respectively the distribution of occupied rapidities for the $n^{\rm th}$ quasiparticle,  the total possible available rapidities for this species (i.e., the density of states), and the occupation function. These quantities are all related to each other through the definition $\vartheta_n(\lambda)=\rho_n(\lambda)/\rho_n^t(\lambda)$ and the Bethe ansatz equations,
\begin{equation}\label{eq:generalBAE}
\rho^t_n(\lambda)=\frac{|p'_n(\lambda)|}{2\pi}-\sum_m\int {\rm d}\mu \,T_{nm}(\lambda-\mu)\vartheta_m(\mu)\rho^t_m(\mu).
\end{equation}
The quasiparticle properties are dependent on the state of the system. In particular the quasiparticle velocity $v_n(\lambda)$ is determined by a similar set of integral equations
\begin{equation}\label{eq:generalVelocity}
v_n(\lambda)\rho^t_n(\lambda)=\frac{\epsilon'_n(\lambda)}{2\pi}-\sum_m\int {\rm d}\mu \,T_{nm}(\lambda-\mu)\vartheta_m(\mu)v_m(\mu)\rho^t_m(\mu)
\end{equation}
while also the effective charge carried by a quasiparticle, $q_{n,{\rm eff}}(\lambda)$ is 
\begin{equation}\label{eq:generalDressedCharge}
q_{n,{\rm eff}}(\lambda)=q_n-\sum_m\int {\rm d}\mu \,T_{nm}(\lambda-\mu)\vartheta_m(\mu)q_{m,{\rm eff}}(\mu).
\end{equation} 
We shall use these expressions to formulate the full time dynamics but before proceeding we briefly comment on our choice of initial states.  

In this work we focus, when discussing integrable models,  on the dynamics emerging from integrable initial states which preserve the integrability of the dynamics in the space direction.  Further to this we take these to be eigenstates of the charge $\hat{Q}$.  In all such cases that we have considered we find that $d_{\alpha,\beta}=d_{\alpha,-\beta}$, i.e., the charged moments in the equilibrium regime are independent of the sign of $\beta$. This can be seen as a consequence of a $\mathbb{Z}_2$ symmetry  whose generator anti-commutes with $\hat{Q}$ and translational invariance.  Consider, for example, the XXZ model quenched from the Majumdar-Gosh or N\'eel states. The Majumdar-Gosh state is an eigenstate of the spin flip operator, $\prod_j\hat{\sigma}^x_j$ which anti-commutes with $S^z$, from which the claim follows.  The N\'eel state on the other hand is not an eigenstate, however the spin flip operator acting on this state is equivalent to translation by a single site and so combining this with the translational invariance of the long time state we also arrive at the same conclusion. Similar considerations can also be applied to other integrable models and initial states. 

As consequence of this $\mathbb{Z}_2$ symmetry we have that $\bar{q}=0$ allowing us to express the charged moments in the following way
\begin{align}\label{eq:spinflipsymmdensity}
\lim_{|A|\to\infty}\lim_{t\to\infty}\log Z_{\alpha,\beta}(A,t)&= |A|\sum_n\int \frac{{\rm d}\lambda}{4\pi}p'_n(\lambda)\left[\mathcal{K}^{(\alpha,\beta)}_n(\lambda)+\mathcal{K}^{(\alpha,-\beta)}_n(\lambda)\right],\\\label{eq:spinflipsymmslope}
\lim_{t\to\infty}\lim_{|A|\to\infty}\log Z_{\alpha,\beta}(A,t)&= 2t\sum_n\int \frac{{\rm d}\lambda}{4\pi}\epsilon'_n(\lambda)\left[\mathcal{L}^{(\alpha,\beta)}_n(\lambda)+\mathcal{L}^{(\alpha,-\beta)}_n(\lambda)\right],
\end{align}
which makes the duality between the two regimes more clear.  We now employ the Bethe Ansatz equations~\eqref{eq:generalBAE} and the velocity equations~\eqref{eq:generalVelocity} to eliminate the bare momentum and energy $p'_n(\lambda),\epsilon'_n(\lambda)$ from these expressions.  Using also~\eqref{eq:Keq} and~\eqref{eq:swappedTBA} of the main text we arrive at
\begin{equation}
  \mkern-10mu
  \begin{aligned}
    \lim_{|A|\to\infty}\lim_{t\to\infty}\log Z_{\alpha,\beta}(A,t) &=|A|\sum_m\int {\rm d}\lambda\, d^{(\alpha,\beta)}_n(\lambda),\\
    &=\frac{1}{2}|A|\sum_{n}\int\mathrm{d}\lambda \,\rho^{t\,(\alpha)}_n(\lambda)
    \left[\mathcal{K}^{(\alpha,\beta)}_n(\lambda)+\mathcal{K}^{(\alpha,-\beta)}_n(\lambda)+\vartheta^{(\alpha)}_n(\lambda)\log[x^{(\alpha,\beta)}_n(\lambda) x^{(\alpha,-\beta)}_n(\lambda)]\right] ,\\
    \lim_{t\to\infty}\lim_{|A|\to\infty}\log Z_{\alpha,\beta}(A,t)&=2t\sum_m\int {\rm d}\lambda\, s^{(\alpha,\beta)}_n(\lambda) \\
    &=t\sum_{n}\int\mathrm{d}\lambda \,\tilde{v}^{(\alpha)}_n(\lambda)\tilde\rho^{t\,(\alpha)}_n(\lambda)
    \left[\mathcal{L}^{(\alpha,\beta)}_n(\lambda)+\mathcal{L}^{(\alpha,-\beta)}_n(\lambda)+\tilde\vartheta^{(\alpha)}_n(\lambda)\log[y^{(\alpha,\beta)}_n(\lambda) y^{(\alpha,-\beta)}_n(\lambda)]\right].\mkern-40mu
  \end{aligned}
  \mkern-20mu
\end{equation}
Here we have introduced filling functions $\vartheta^{(\alpha)}_n(\lambda)$ and
$\tilde{\vartheta}^{(\alpha)}_n(\lambda)$ as
\begin{equation}
\vartheta^{(\alpha)}_n(\lambda)=
  \frac{\vartheta_n^\alpha e^{-\text{sgn}[p'_n]\log x_n^{(\alpha,0)}}}{(1-\vartheta_n)^\alpha+\vartheta_n^\alpha e^{-\text{sgn}[p'_n]\log x_n^{(\alpha,0)}}},
  \qquad
\tilde\vartheta^{(\alpha)}_n(\lambda)=
  \frac{\vartheta_n^\alpha e^{-\text{sgn}[\epsilon'_n]\log y_n^{(\alpha,0)}}}{(1-\vartheta_n)^\alpha+\vartheta_n^\alpha e^{-\text{sgn}[\epsilon'_n]\log y_n^{(\alpha,0)}}},
\end{equation}
with $x^{(\alpha,\beta)}_n$ and $y^{(\alpha,\beta)}_n$ solutions to
Eqs.~\eqref{eq:Keq} and~\eqref{eq:swappedTBA} of the main text, while
$\rho^{t,(\alpha)}_n$, $\rho^{(\alpha)}_n$, $\tilde{\rho}^{t,(\alpha)}_n$,
$\tilde{\rho}^{(\alpha)}_n$, and $\tilde{v}^{(\alpha)}_n$ are the associated
rapidity distributions and quasiparticle velocity.

We now finally arrive at the expression for the full time dynamics (cf.\ Eq.~\eqref{eq:fullconj} of the main text),
\begin{equation}
  \begin{aligned}
    \log Z_{\alpha,\beta}(A,t)&=\sum_n\int{\rm d}\lambda\, 
    \text{min}(|A|d^{(\alpha,\beta)}_n(\lambda),2t s^{(\alpha,\beta)}_n(\lambda))\\
    &=\sum_n\int{\rm d}\lambda \,
    \text{min}(|A|,2t v^{(\alpha,\beta)}_n(\lambda))d^{(\alpha,\beta)}_n(\lambda)
  \end{aligned}
\end{equation}
where in the second line we have defined $v^{(\alpha,\beta)}_n(\lambda)=s_n^{(\alpha,\beta)}(\lambda)/d^{(\alpha,\beta)}_n(\lambda)$. We note that the choice of $\vartheta^{(\alpha)}_n(\lambda)$ and $\tilde\vartheta^{(\alpha)}_n(\lambda)$ are somewhat arbitrary as any stationary state could be used to eliminate $p'_n, \epsilon_n'$ from~\eqref{eq:spinflipsymmdensity} and ~\eqref{eq:spinflipsymmslope}, however, our choice allows us to reproduce the quasiparticle-picture result for the entanglement entropy~\cite{alba2017entanglement}. Moreover, as shown in Sec.~\ref{sec:FTCMs}, in the limit $\alpha\to1$ this choice also reproduces the quasiparticle expression for the equal-time connected correlation function of the charge density.  

\subsection{Fourier Transform of the Charge moments}\label{sec:FTCMs}
Armed with the expression for the full time dynamics we now look to compute 
\begin{equation}
\tr[\rho_{A,Q}^\alpha(t)]=\int_{-\pi}^\pi \frac{\rm{d}\beta}{2\pi}Z_{\alpha,\beta}(A,t)e^{-i\beta Q},
\end{equation}
which is necessary for determining the SREE and also the probability distribution of the charge (cf.\ Eq.~\eqref{eq:lastEq} of the main text). To achieve this we note that for $t,|A|\gg 1$ the exponent of $Z_{\alpha,\beta}(A,t)$ is rapidly oscillating and  we can compute the integral via a stationary phase approximation. In addition by considering $Q/|A|$ to be small we can proceed as in the Sec.~\ref{sec:details} and find
\begin{equation}
  \tr[\rho_{A,Q}^\alpha(t)]\simeq
  \frac{1}{\sqrt{2\pi\mathcal{D}_{\alpha}(t) }}Z_{\alpha}(A,t)e^{-\frac{Q^2}{2\mathcal{D}_{\alpha}(t)}},
\end{equation}
where 
\begin{equation}
  \mathcal{D}_{\alpha}(t)=-\partial_{\beta}^2\log Z_{\alpha,\beta}(A,t)|_{\beta=0}.
\end{equation}
To evaluate this we use that $\partial_\beta \log x^{(\alpha,\beta)}_n(\lambda)|_{\beta=0}=-i q^{(\alpha)}_{n,{\rm eff}}(\lambda)$ where $q^{(\alpha)}_{n,{\rm eff}}(\lambda)$ is the effective quasiparticle charge in the state specified by $\vartheta^{(\alpha)}_n$.  While also $\partial_\beta \log y^{(\alpha,\beta)}_n(\lambda)|_{\beta=0}=-i \tilde{q}^{(\alpha)}_{n,{\rm eff}}(\lambda)$ where $\tilde{q}^{(\alpha)}_{n,{\rm eff}}(\lambda)$ is the effective quasiparticle charge in the state specified by $\tilde{\vartheta}^{(\alpha)}_n$.  Then since min$(x,y)=x\,\Theta(y-x)+y\,\Theta(x-y)$,  with $\Theta(z)$ being the Heaviside function,  and the fact that the first derivative at $\beta=0$ vanishes we find
\begin{equation}
  \begin{split}
    \mathcal{D}_{\alpha}(t)=\sum_n\int {\rm d}  \lambda\, |A| (q^{(\alpha)}_{n,{\rm eff}})^2\rho^{(\alpha)}_n(1-\vartheta^{(\alpha)}_n)\Theta(2ts^{(\alpha\,0)}_n-|A|d^{(\alpha\,0)}_n)\\
    +
    \sum_n\int {\rm d}  \lambda\, 2 t \tilde{v}^{(\alpha)}_n(\tilde q^{(\alpha)}_{n,{\rm eff}})^2\tilde \rho^{(\alpha)}_n(1-\tilde \vartheta^{(\alpha)}_n)\Theta(|A|d^{(\alpha\,0)}_n-2ts^{(\alpha\,0)}_n),
  \end{split}
\end{equation}
which evidently shows a breakdown of the quasiparticle picture due the $\alpha$ dependent dressing of the quantities involved. 

An important quantity from this family is $\alpha=1$ which provides the probability distribution of the charge.  To take this limit we need
\begin{equation}
  \begin{aligned}
\lim_{\alpha\to 1}\Theta(2ts^{(\alpha\,0)}_n(\lambda)-|A|d_n^{(\alpha\,0)}(\lambda))&=
  \lim_{\alpha\to 1}\Theta\left(\frac{1}{\alpha-1}
  \left[2ts^{(\alpha\,0)}_n(\lambda)-|A|d_n^{(\alpha\,0)}(\lambda)\right]\right)\\
    &=\Theta(2t|v_n(\lambda)|-|A|),
  \end{aligned}
\end{equation}
where we have used $\lim_{\alpha\to 1}\vartheta^{(\alpha)}_n,\tilde\vartheta^{(\alpha)}_n=\vartheta_n$. It follows then that $\mathcal{D}(t)=\lim_{\alpha\to 1}\mathcal{D}^{(\alpha)}(t)$ is 
\begin{equation}
\mathcal{D}(t)=\sum_n\int {\rm d}\lambda \, q_{n,{\rm eff}}^2\rho_n(1-\vartheta_n)\text{min}(2t |v_n(\lambda)|,|A|)
\end{equation}
with $q_{n,{\rm eff}}=\lim_{\alpha\to 1}q^{(\alpha)}_{n,{\rm eff}}$.  Therefore, analogously to the case of the von Neuman entanglement entropy, a quasiparticle picture reemerges at $\alpha\to 1$.

\section{Equivalence between the TBA prediction and exact result for Rule 54}\label{sec:Rule54}

The conjecture~\eqref{eq:swappeddensity} can be verified exactly in the case of
the so-called \emph{Rule 54} quantum cellular automaton~\cite{bobenko1993two}.  This model is interacting, Yang-Baxter integrable~\cite{gombor2022integrable}, and treatable by TBA~\cite{friedman2019integrable}, however, in contrast with generic interacting integrable models, it allows for the exact calculation of several nonequilibrium quantities~\cite{prosen2016integrability,prosen2017exact,gopalakrishnan2018operator,gopalakrishnan2018hydrodynamics,inoue2018two,friedman2019integrable,alba2019operator,klobas2019time,buca2019exact,alba2021diffusion,klobas2020matrix,klobas2020space,klobas2021exact,klobas2021exactrelaxation,klobas2021entanglement} (see Ref.~\cite{buca2021rule} for a recent review). For this reason it can be considered the \emph{minimal} interacting integrable model. 

Rule 54 is formulated as a quantum circuit of qubits (${d=2}$) with the evolution
implemented by 
\begin{equation}
\hat{\mathbb{U}}=\hat{\Pi}^{\dagger} \hat{\mathbb{U}}_{\mathrm{e}} 
  \hat{\Pi} \hat{\mathbb{U}}_{\mathrm{e}},\qquad
\hat{\mathbb{U}}_{\mathrm{e}}= \prod_{j\in \mathbb{Z}} \hat{U}_{2j}.
\end{equation}
Here $\hat{U}_j$ acts non-trivially only on three sites ($j-1$, $j$, and $j+1$) implementing the following deterministic update of the middle one
\begin{equation}
\mel{s_1^{\prime} s_2^{\prime} s_3^{\prime}}{\hat{U}}{s_1^{\phantom{\prime}} s_2^{\phantom{\prime}} s_3^{\phantom{\prime}}}
  =
  \delta_{s_1^{{\prime}},s_1}
  \delta_{s_2^{{\prime}},\chi(s_1,s_2,s_3)}
  \delta_{s_3^{{\prime}},s_3},\qquad
  \chi(s_1,s_2,s_3)\equiv s_1+s_2+s_3+s_1 s_3\pmod{2}.
\end{equation}
The time-evolution can be alternatively represented in terms of a tensor network with local interactions~\cite{klobas2021exact}, which one can use to evaluate the CM following the reasoning outlined above for brickwork quantum circuits~\cite{klobas2023inprep}. In particular, considering a quench from a \emph{solvable} initial state~\cite{klobas2021exactrelaxation} $\ket*{\Psi_0}=(\sqrt{1-\vartheta}\ket{00}+\sqrt{\vartheta}\ket{01})^{\otimes L/2}$ parametrized by a parameter $0<\vartheta<1$ that fixes the quasiparticle occupations, and focussing on the conserved quantity
$\hat{Q}^{(-)}=\sum_{j} (-1)^j \sigma_{j}^z \sigma_{j+1}^z$ one obtains the following exact result~\cite{klobas2023inprep}
\begin{equation}\label{eq:ExactResultRule54a}
  \lim_{t\to\infty}
  \lim_{\abs{A}\to\infty}
  \lim_{\bar{\abs{A}}\to\infty}
  \frac{1}{t}\log Z_{\alpha,\beta}(A,t)=
  s_{\alpha,\beta}+s_{\alpha,-\beta},
\end{equation}
where $s_{\alpha,\beta}$ satisfies
\begin{equation} \label{eq:ExactResultRule54b}
  s_{\alpha,\beta}=
  2\log[(1-\vartheta)^{\alpha}
  +\frac{e^{i\beta}\vartheta^{\alpha}}{e^{s_{\alpha,\beta}}}].
\end{equation}
This expression is immediately reproduced by taking the
conjecture~\eqref{eq:swappeddensity} and~\eqref{eq:swappedTBA}, and plugging in
the TBA data for Rule 54. To see this explicitly we start by recalling the
TBA formulation of the model (see e.g.~\cite{friedman2019integrable}), which
consists of two particle species, labelled by $n\in\{+1,-1\}$, and
\begin{equation}
    \vartheta_n(\lambda)=\vartheta_n,\qquad
    \varepsilon^{\prime}_n(\lambda)=\tilde{q}_n=n,\qquad
    p^{\prime}_n(\lambda)=1,\qquad
    \int\frac{\mathrm{d}\mu}{2\pi}T_{nm}(\lambda-\mu)=n m.
\end{equation}
Using this identification, we obtain the following simple form of
Eqs.~\eqref{eq:swappeddensity} and~\eqref{eq:swappedTBA},
\begin{equation}
    \log y_{+}=-\log y_{-}=s_{\alpha,\beta}+i\beta
    = i\beta
    +\log[(1+\vartheta_+)^{\alpha}+\frac{\vartheta_{+}^{\alpha}}{y_{+}}]
    +\log[(1+\vartheta_-)^{\alpha}+\frac{\vartheta_{-}^{\alpha}}{y_{+}}].
\end{equation}
Next, we note that the stationary state after the quench from
$\ket{\Psi_0(\vartheta)}$ is parametrized by
$(\vartheta_{+},\vartheta_{-})=(\vartheta,\vartheta)$~\cite{klobas2021exactrelaxation},
which directly gives Eq.~\eqref{eq:ExactResultRule54b}. Finally, to get the
expression~\eqref{eq:ExactResultRule54a}, we remark that
$\hat{Q}^{-}\ket*{\Psi_0}=0$, and hence there is no dependence on the size of
the subsystem (cf.\ Eq.~\eqref{eq:chargemomentprediction}).

\section{The TBA formulation of the conjecture for the XXZ model at principal roots of unity}\label{sec:XXZBethe}
\subsection{The TBA description of XXZ at roots of unity}

Here we test our prediction for the XXZ spin-$1/2$ chain: the paradigmatic example of interacting integrable model. This is a lattice system with continuous time evolution generated by the Hamiltonian
\be
\hat H=\sum_{j=1}^{2L} \hat \sigma^x_{j}\hat\sigma^x_{j+1}+ \hat\sigma^y_{j}\hat\sigma^y_{j+1}+ \Delta \hat\sigma^z_{j}\hat\sigma^z_{j+1},
\ee
where $\hat\sigma^{x,y,z}_j,$ are Pauli matrices, the parameter 
$\Delta\in \mathbb R$ is referred-to as anisotropy, and we consider periodic boundary conditions.  For definiteness we restrict ourselves to $|\Delta|\leq 1$ 
and parameterise the anisotropy parameter $\Delta$
 as 
\begin{equation}
  \Delta=\cos\gamma,\qquad \gamma=\frac{\pi}{p+1},
\end{equation}
with $p$ being a positive integer. Other regimes can be considered similarly and will be presented
elsewhere~\cite{bertini2023dynamics}.

 We take ${\hat Q=\hat S^z}$, the $z$ component of the magnetisation,  while the initial state is chosen to be either the N\'eel state, 
 $$\ket*{\Psi_{\rm{N}}}=\ket{\uparrow\downarrow}^{\otimes L},$$
 or the Majumdar-Gosh state, 
 $$\ket*{\Psi_{\rm{MG}}}=\left[\frac{1}{\sqrt{2}}\left(\ket{\uparrow\downarrow}-\ket{\downarrow\uparrow}\right)\right]^{\otimes L}.$$
 These are ``integrable'' initial states for which $\vartheta_n(\lambda)$ can be computed analytically~\cite{ilievski2016string, piroli2017what} which is discussed further below.  

For the choice of parameters under consideration the spectrum of the system consists of $p+1$ families of stable quasiparticles.  With these specifications and for $p=1$ we reproduce the exact noninteracting result for the XX chain~\cite{parez2021exact}, while for ${\beta=0}$ and ${p>1}$ we obtain the dynamics of the Reny\'{i} entropies in the interacting model~\cite{bertini2022growth}. To check our results for ${\beta\neq0}$, ${p>1}$ we compare Eqs.~\eqref{eq:swappeddensity} and~\eqref{eq:swappedTBA} against numerical simulations using Matrix Product State (MPS) based algorithms~\cite{vidal2003mps}. In particular we use infinite time-evolving block decimation scheme (iTEBD)~\cite{vidal2007itebd} that, exploiting two-site shift invariance of $\ket{\Psi_t}$, works directly in the thermodynamic limit.  

Below we report further details on the numerical techniques while here we
report all the equations needed to evaluate evaluate
Eqs.~\eqref{eq:deq}-\eqref{eq:swappedTBA} for the XXZ. As mentioned above in
this case the number of quaisparticle species (also known as strings) is $p+1$,
therefore the summation index $n$ runs between $1$, and $p+1$. We introduce
the string parity $\nu_n$ and the string length $q_n$ (equivalent to the charge
of the string), and  are given by
\begin{equation}
  \nu_{n}=
  \begin{cases}
    1,& n\le p,\\ -1,& n=p+1,
  \end{cases},\qquad
  q_n=
  \begin{cases}
    n,& n\le p,\\ 1,& n=p+1.
  \end{cases}
\end{equation}
The rapidities take values on the full real line, and for convenience we define
the convolution as
\begin{equation}
  (f\star g)(\lambda)=\int_{-\infty}^{\infty} {\rm d}\mu f(\lambda-\mu)g(\mu).
\end{equation}
As stated in the main text, the system is described through its quasiparticle
content which have known expressions for their energy and momentum in terms of
their rapidity $\lambda$ and species index $j$.  In the thermodynamic limit we
describe the system through the distributions of these rapidites which we
denote $\rho_j(\lambda)$ as well as the distributions for the unoccupied
rapidities $\rho_j^h(\lambda)$. The derivatives of the energy and momentum with
respect to the rapidity as well as the scattering kernel for the model can be
given through the family of functions $\tilde{a}_j^{\nu}(\lambda)$ defined for
$j=1,\ldots p$, and $\nu=\pm1$,
\begin{equation}
  \tilde{a}_{j}^{\nu}(\lambda)=
  \frac{\nu\sin\frac{j\pi}{p+1}}{\pi\left(\cosh(2\lambda)-\nu\cos\frac{j \pi}{p+1}\right)},
\end{equation}
in terms of which we have 
\begin{equation} \label{eq:definitionsTBAgapless}
  \epsilon^{\prime}_j(\lambda)=-\pi \sin(\gamma)a'_j(\lambda),\qquad
  p'_j(\lambda)=2 \pi a_j(\lambda), \qquad a_j(\lambda)=\tilde{a}_{q_j}^{\nu_j}(\lambda),
\end{equation}
and
\begin{equation}
    T_{jk}(\lambda)=
    (1-\delta_{q_j,q_k})\tilde{a}_{\abs*{q_j-q_k}}^{\nu_j\nu_k}(\lambda)
    +2\left(\tilde{a}_{\abs*{q_j-q_k}+2}^{\nu_j\nu_k}(\lambda)
    +\tilde{a}_{\abs*{q_j-q_k}+4}^{\nu_j\nu_k}(\lambda)
    +\ldots
    +\tilde{a}_{q_j+q_k-2}^{\nu_j\nu_k}(\lambda)\right)
    +\tilde{a}_{q_j+q_k}^{\nu_j\nu_k}(\lambda).
\end{equation}
For convenience we define the boundary values as
\begin{equation}
  a_0(\lambda)=\delta(\lambda),\qquad T_{0k}(\lambda)=T_{k0}(\lambda)=0.
\end{equation}
With this definition, the functions $a_{j}(\lambda)$, $T_{kl}(\lambda)$ can be
shown to satisfy the following sets of relations,
\begin{equation}\label{eq:relationsAandT}
  \begin{aligned}
    1\le j \le p-1:& &
    T_{jk}&=s\star(T_{j-1k}+T_{j+1k})
    +(\delta_{j-1k}+\delta_{j+1k}-\delta_{j,p-1}\delta_{k,p+1}) s, 
    &\qquad a_j&=s\star(a_{j-1}+a_{j+1})\\
    & & T_{pk}&=-T_{p+1k}=s\star T_{p-1k} + \delta_{p-1k} s, 
    & a_p &=-a_{p+1}=s\star a_{p-1},
  \end{aligned}
\end{equation}
with the function $s(\lambda)$ given as
\begin{equation}
  s(\lambda)=\frac{p+1}{2\pi\cosh\left((p+1)\lambda\right)}.
\end{equation}
We note that the form of the relation for $T_{jk}(\lambda)$ in 
Ref.~\cite{takahashi1999thermodynamics} contains a misprint that is corrected here.

The rapidity distributions $\rho_j$, and the quasiparticle velocities $v_j$ --- together with
$\rho^t_j=\rho_j+\rho^h_j$, and $v_j\rho^t_j$ --- satisfy the following set of coupled integral
equations,
\begin{equation}\label{eq:rhoTrhoTv}
  \rho^{t}_{j}=\nu_j a_j 
  - \nu_j \sum_{k=1}^{p+1} T_{jk}\star \rho_k,\qquad
  v_j \rho^{t}_{j}=-\frac{\nu_j\sin\gamma}{2}a_j^{\prime}
  - \nu_j \sum_{k=1}^{p+1} T_{jk}\star (v_k \rho_k).
\end{equation}
Upon using ~\eqref{eq:relationsAandT} these can be written in  the following decoupled form
\begin{equation}\label{eq:rhoTrhoTvDecoupled}
  \begin{aligned}
    \rho^{t}_1&=s + s\star\left(\rho_{2}^h+\delta_{p,2} \rho_{p+1}\right),&
    \rho^{t}_1 v_1 & = -\frac{\sin\gamma}{2} s^{\prime}
    +s \star\left(\rho_2^h v_2 +\delta_{p,2}\rho_{p+1}v_{p+1}\right),\\
    1<j<p:\ 
    \rho^t_j &=s\star\left(\rho_{j-1}^h+\rho_{j+1}^h+\delta_{p,j+1}\rho_{p+1}\right),&
    \rho^t_jv_j &=s\star\left(\rho_{j-1}^hv_{j-1}+\rho_{j+1}^h v_{j+1}
    +\delta_{p,j+1}\rho_{p+1}v_{p+1}\right),\\
    \rho^t_p &=\rho^t_{p+1}=s\star \rho_{p-1}^h,&
    \rho^t_p v_p &=\rho^t_{p+1} v_{p+1}=s\star \rho_{p-1}^h v_{p-1}.
  \end{aligned}
\end{equation}
The filling function $\vartheta_j(\lambda)=(1+\eta_j(\lambda))^{-1}$ depends on
the initial state, and the defining equations for the cases considered here are
given in Appendix~\ref{sec:etas}.

\subsection{The TBA form of the conjecture}
In the main text we have implicitly assumed that the string parity $\nu_n$ is
always positive. However, whenever this does not hold (as is the case here),
the correct result is recovered by performing the replacement
\begin{equation}
  T_{nm}(\lambda)\mapsto \nu_n T_{nm}(\lambda),\qquad q_{n}\mapsto \nu_n q_n,\qquad
  \tilde{q}_n\mapsto \nu_n\tilde{q}_n,
\end{equation}
which in particular implies that the defining equations for auxiliary functions
$\log x_{n}$ and $\log y_n$ have to be modified (cf.\ Eqs.~\eqref{eq:Keq},
and~\eqref{eq:swappedTBA}),
\begin{equation}
    \log[x_{n}] = i \beta \nu_n q_n + \sum_{m} (\nu_n T_{nm}\star \mathcal{K}_{m}^{(\alpha,\beta)}),\qquad
    \log[y_{n}] = i \beta \nu_n \tilde{q}_n + \sum_{m} (\nu_n T_{nm}\star \mathcal{L}_{m}^{(\alpha,\beta)}).
\end{equation}
This can be specialized to our case when we take into account
\begin{equation}
  \tilde{q}_n=q_n,\qquad \mathrm{sgn}(p^{\prime}_n(\lambda))=\nu_n,\qquad
  \mathrm{sgn}(\varepsilon^{\prime}_n(\lambda))=\nu_n \mathrm{sgn}(\lambda),
\end{equation}
which implies the following form of Eqs.~\eqref{eq:Keq} and~\eqref{eq:swappedTBA},
\begin{equation} \label{eq:equationsGapless}
  \begin{aligned}
    \mathcal{K}_{n}^{(\alpha,\beta)}&=
    \nu_{n} \log[(1-\vartheta_n)^{\alpha}
    +\frac{\vartheta_n^{\alpha}}{x_n^{\nu}}],&\qquad
    \mathcal{L}_{n}^{(\alpha,\beta)}&=
    \mathrm{sgn}(\cdot)\nu_n \log[(1-\vartheta_n)^{\alpha}
    +\frac{\vartheta_n^{\alpha}}{y_n^{\mathrm{sgn}(\cdot) \nu_n}}],\\
    \log[x_{n}] &= 
    i \beta \nu_n q_n + \sum_{m} (\nu_n T_{nm}\star \mathcal{K}_{m}^{(\alpha,\beta)}),&
    \log[y_{n}] &= 
    i \beta \nu_n q_n + \sum_{m} (\nu_n T_{nm}\star \mathcal{L}_{m}^{(\alpha,\beta)}).
  \end{aligned}
\end{equation}

Now we introduce the following shorthand notation,
\begin{equation}
  \begin{aligned}
    X_{n}^{(\alpha,\beta)}&=\log x_{n}-i\beta \nu_n q_n, & 
    Y_{n}^{(\alpha,\beta)}&=\log y_{n}-i\beta \nu_n q_n,\\
    \tilde{\mathcal{K}}_{n}^{(\alpha,\beta)}&
    =\mathcal{K}_n^{(\alpha,\beta)}+X^{(\alpha,\beta)}_{n}, &
    \tilde{\mathcal{L}}_{n}^{(\alpha,\beta)}&
    =\mathcal{L}_n^{(\alpha,\beta)}+Y^{(\alpha,\beta)}_{n},
  \end{aligned}
\end{equation}
which allows us to rewrite the bottom line of~\eqref{eq:equationsGapless} as
\begin{equation}
  X_{n}^{(\alpha,\beta)}=\sum_{m}\nu_n T_{n,m}\star \mathcal{K}_{m}^{(\alpha,\beta)},\qquad
  Y_{n}^{(\alpha,\beta)}=\sum_{m}\nu_n T_{n,m}\star \mathcal{L}_{m}^{(\alpha,\beta)},
\end{equation}
which can be, using the relations~\eqref{eq:relationsAandT}, recast in a decoupled form,
\begin{equation}
  \begin{aligned}
    X_1^{(\alpha,\beta)}&=s\star(\tilde{\mathcal{K}}^{(\alpha,\beta)}_2
    -\delta_{p,2} \mathcal{K}^{(\alpha,\beta)}_{p+1}),\quad&
    Y_1^{(\alpha,\beta)}&=s\star(\tilde{\mathcal{L}}^{(\alpha,\beta)}_2
    -\delta_{p,2} \mathcal{L}^{(\alpha,\beta)}_{p+1}),\\
    1<n<p:\ X_n^{(\alpha,\beta)}&=s\star(
    \tilde{\mathcal{K}}^{(\alpha,\beta)}_{n-1}+ \tilde{\mathcal{K}}^{(\alpha,\beta)}_{n+1}
    -\delta_{p,n+1} \mathcal{K}^{(\alpha,\beta)}_{p+1}),&
    Y_n^{(\alpha,\beta)}&=s\star(
    \tilde{\mathcal{L}}^{(\alpha,\beta)}_{n-1}+ \tilde{\mathcal{L}}^{(\alpha,\beta)}_{n+1}
    -\delta_{p,n+1} \mathcal{L}^{(\alpha,\beta)}_{p+1}),\\
    X_{p}^{(\alpha,\beta)}&=X_{p+1}^{(\alpha,\beta)}=s\star \tilde{\mathcal{K}}^{(\alpha,\beta)}_{p-1},&
    Y_{p}^{(\alpha,\beta)}&=Y_{p+1}^{(\alpha,\beta)}=s\star \tilde{\mathcal{L}}^{(\alpha,\beta)}_{p-1}.
  \end{aligned}
\end{equation}
Having established this convenient notation we can express the asymptotic
behaviour of charged moments in both regimes in an equivalent form that behaves better in
numerical manipulations. First the equilibrium density,
\begin{equation}
  \begin{aligned}
    d_{\alpha,\beta}= \frac{i\beta}{2}+
    \sum_{n} \int\!\mathrm{d}\lambda a_n(\lambda) \mathcal{K}_n^{(\alpha,\beta)}(\lambda)
    =\sum_{n}\nu_n
    \int\!\mathrm{d}\lambda \rho_n^{t}(\lambda)\left(\mathcal{K}_n^{(\alpha,\beta)}(\lambda)
    +\vartheta_n(\lambda)\log x_{n}(\lambda)\right),
  \end{aligned}
\end{equation}
where in the first equality we took into account $q_0=1/2$, and the definition
of $p^{\prime}_n$~\eqref{eq:definitionsTBAgapless}, and in the second the
equation for $\rho_t$~\eqref{eq:rhoTrhoTv}, and 
\begin{equation}
  \sum_{n} q_{n}\int\!\mathrm{d}\lambda \rho_{n}(\lambda)=q_0=\frac{1}{2}.
\end{equation}
Similarly we express the contribution to the asymptotic slope $s_{\alpha,\beta}$,
\begin{equation}
  s_{\alpha,\beta}=\sum_n\int\!\mathrm{d}\lambda
  \left(-\frac{\sin\gamma}{2} a_n^{\prime}\right) \mathcal{L}_n^{(\alpha,\beta)}(\lambda)
  =\sum_n\nu_n\int\!\mathrm{d}\lambda
  v_n(\lambda)\rho^t_n(\lambda)\left(\mathcal{L}_n^{(\alpha,\beta)}(\lambda)
  +\vartheta_n(\lambda)\log y_n(\lambda)\right).
\end{equation}
Note that here we have $\tilde{q}_0=0$, which is satisfied both in the case of
N\'eel and dimer state. Moreover, we also used that $v_{n}(\lambda)$, and $\rho_n(\lambda)$
are respectively an odd and even function of $\lambda$, which for all $n$ implies
\begin{equation}
  \int\, \mathrm{d}\lambda \rho_n(\lambda) v_n(\lambda)=0.
\end{equation}

\subsection{Saddle-point equation for \texorpdfstring{$\eta_j$}{etas}}\label{sec:etas}
The first initial state we consider is the \emph{N\'eel state},
\begin{equation}
  \ket*{\Psi_0^{(\rm N)}}=\ket{\uparrow\downarrow}^{\otimes L/2},
\end{equation}
for which the saddle point equations for $\eta_j=\vartheta_j^{-1}-1$ read as,
\begin{equation}
  \begin{aligned}
    & & \log {\eta_1}&=(1+\delta_{p,2})s\star \log(1+\eta_2) -d(\lambda),\\
    1<j<p:& &
    \log {\eta_j}&=s\star\left[\log(1+\eta_{j-1})+(1+\delta_{p,j+1}) \log(1+\eta_{j+1})\right]
    +(-1)^j d(\lambda),\\
    p \text{ even}:& &
    \log {\eta_p}&=s\star\log(1+\eta_{p-1})+d(\lambda),\\
    p \text{ odd}:& &
    \log \eta_p&=s\star\log(1+\eta_{p-1}),\\
    & & \log \eta_{p+1}&=-\log\eta_{p},
  \end{aligned}
\end{equation}
with the function $d(\lambda)$ taking the following form,
\begin{equation}
  d(\lambda)=\log\Big[\Big(\coth\frac{(p+1)\lambda}{2}\Big)^2\Big].
\end{equation}
The second initial state is the \textit{Majumdar-Gosh} state,
\begin{equation}
\ket*{\Psi_0^{(\rm{MG})}}=\left[\frac{1}{\sqrt{2}}\left(\ket{\uparrow\downarrow}-\ket{\downarrow\uparrow}\right)\right]^{\otimes L/2},
\end{equation}
for which we have a similar set of equations,
\begin{equation}
  \begin{aligned}
 \log {\eta_1}&=(1+\delta_{p,2})s\star \log(1+\eta_2) -d(\lambda),\\
    \log {\eta_j}&=s\star\left[\log(1+\eta_{j-1})+(1+\delta_{p,j+1}) \log(1+\eta_{j+1})\right]
    -d(\lambda),\\
    \log {\eta_p}&=s\star\log(1+\eta_{p-1})-d(\lambda),\\
 \log \eta_{p+1}&=-\log\eta_{p}.
  \end{aligned}
\end{equation}

\subsection{Fourier transforms of relevant functions}
We conclude this appendix by listing the Fourier transforms of some of the
functions introduced here that might be of use in the numerical implementation
of TBA equations. First we fix the conventions for the Fourier transform and the
inverse Fourier transform as,
\begin{equation}
  \hat{f}(k)=\int_{-\infty}^{\infty} \mathrm{d}x e^{i k x},\qquad
  f(x)=\frac{1}{2\pi}\int_{-\infty}^{\infty}\mathrm{d} k e^{-i k x},
\end{equation}
which implies
\begin{equation}
  \hat{(f\star g)}(k)= \hat{f}(k)\hat{g}(k).
\end{equation}
With this convention we obtain the following:
\begin{equation}
  \begin{gathered}
    \hat{s}(k)=\frac{1}{2\cosh\frac{k\pi}{2(p+1)}},\qquad
    \hat{d}(k)=\frac{2 \pi}{k} \tanh\frac{k\pi}{2(p+1)},\qquad
    \hat{s^{\prime}}=
    \frac{-i k}{2\cosh\frac{k\pi}{2(p+1)}},\\
    \hat{\tilde{a}}^{+1}_n(k)=\frac{\sinh\frac{k\pi (p+1-n)}{2(p+1)}}{\sinh\frac{k\pi}{2}},
    \qquad \hat{\tilde{a}}^{+1}_0(k)=1,\qquad
    \hat{\tilde{a}}^{-1}_n(k)=-\frac{\sinh\frac{k\pi n}{2(p+1)}}{\sinh\frac{k\pi}{2}}.
  \end{gathered}
\end{equation}

\section{Further details about the numerical calculations}
\label{sec:Numerics}

Numerical simulations of the quench dynamics in the XXZ spin-1/2 chain have been performed via Tensor-Network based algorithms. The many-body wave function has been represented as a two-site shift invariant Matrix Product State (MPS) as follows (in the following we discard the explicit time-dependence in all matrices and tensors to simplify the notation)
\be\label{eq:psi_mps}
|\Psi_t\rangle = 
(v_l|
\cdots
{\bf \Gamma}_e
{\bf \Lambda}_e
{\bf \Gamma}_o
{\bf \Lambda}_o
\cdots
|v_r),
\ee
where we introduced the vector-valued matrices 
${\bf \Gamma}_{e/o} = \sum_{\sigma\in\{+1,-1\}}{\bf \Gamma}^{\sigma}_{e/o} |\sigma\rangle$, and the diagonal matrices ${\bf \Lambda}_{o/e}$ with real entries. All matrices have bond dimensions $\chi$.

Indeed, thanks to a number of MPS symmetries, 
left and right boundary vectors have been chosen 
such that $ |v_r)(v_r| = |v_l)(v_l| = \mathbb{I}$, and
the following canonical equations apply 
\be
\sum_{\sigma}{\bf \Gamma}^{\sigma}_{e/o}{\bf \Lambda}^{2}_{e/o} {\bf \Gamma}^{\sigma\dag}_{e/o} = \mathbb{I}, \quad
\sum_{\sigma}{\bf \Gamma}^{\sigma\dag}_{e/o}{\bf \Lambda}^{2}_{o/e} {\bf \Gamma}^{\sigma}_{e/o} = \mathbb{I}.
\ee
Thanks to this fact, the diagonal entries of the matrices ${\bf\Lambda}_{e/o}$ do represent the Schmidt coefficient associated to the Schmidt decomposition of the state $|\Psi_t\rangle$ across any even/odd bond of the chain.

The representation in Eq.~(\ref{eq:psi_mps}) holds true at each instant of time, provided that the evolution operator ${\rm exp}(-i t H)$ is approximated via a Suzuki-Trotter expansion in a checkerboard fashion.
In particular, in our time evolution, we use a second-order expansion such that
\be\label{eq:U_trotter}
{\rm exp}(-i dt H) = 
\prod_{j\;{\rm odd}} e^{- i dt h_j /2}
\prod_{j\;{\rm even}} e^{- i dt h_j}
\prod_{j\;{\rm odd}} e^{- i dt h_j /2} + O(dt^3),
\ee
with Hamiltonian density 
$h_{j} = s^{x}_{j}s^{x}_{j+1}+ s^{y}_{j}s^{y}_{j+1} + \Delta s^{z}_{j}s^{z}_{j+1}$ and Trotter time-step $dt = 10^{-2}$.
By using the iTEBD algorithm~\cite{vidal2007itebd}, we can easily apply the local operators in Eq.~(\ref{eq:U_trotter}) to the state $|\Psi_t\rangle$, and finally recast $|\Psi_{t+dt}\rangle = e^{-i dt H}|\Psi_{t}\rangle $ in the same canonical form of a two-site shift invariant MPS.  
In our simulations we let the auxiliary dimension grow up to $\chi_{max} = 1024$, being able to reach $t_{max}=12$ without appreciable numerical error.

At any time, the reduced density matrix of a semi-infinite chain has therefore the following diagonal representation (assuming we divide the system across an even bond)
\be
\rho_{[0,\infty]} = \sum_{k=1}^{\chi} [{\bf \Lambda}^{2}_{e}]_{kk} |\phi_{k}\rangle\langle \phi_{k}|
\ee
with Schmidt vectors $|\phi_{k}\rangle = 
\sum_{\sigma_1,\sigma_2,\dots} \left[
{\bf \Gamma}^{\sigma_{1}}_o{\bf \Lambda}_o
{\bf \Gamma}^{\sigma_{2}}_e{\bf \Lambda}_e
\cdots
|v_r) \right]_{k1} |\sigma_{1},\sigma_{2},\dots\rangle$,
such that $\langle \phi_{p}|\phi_{q}\rangle = \delta_{pq}$.
From that, we easily computed  \be F_{\alpha,\beta}(A,t) = {\rm Tr} \left[ \rho_{[0,\infty]}(t)^\alpha e^{i \beta S^{z}_A})\right], \ee up to time $t_{max} = 12$.
As a matter of fact, the expectation value of the generating function of the subsystem magnetization
$
e^{i \beta S^{z}_{A}} = \prod_{j=1}^{\ell} e^{i \beta s^{z}_{j}}
$
can be easily computed, for any power of the reduced density matrix $\rho^{\alpha}_{[0,\infty]}$, as
\be
F_{\alpha,\beta}(A,t)
= 
{\rm Tr} \left[
\sum_{\sigma_{\ell}} e^{i \beta \sigma_{\ell}/2}
{\bf \Lambda}_{e/o} {\bf \Gamma}^{\dag\sigma_{\ell}}_{e/o}
\cdots
\left(
\sum_{\sigma_{2}} e^{i \beta \sigma_{2}/2}
{\bf \Lambda}_{e} {\bf \Gamma}^{\dag\sigma_2}_{e}
\left(
\sum_{\sigma_{1}} e^{i \beta \sigma_{1}/2}
{\bf \Lambda}_{o} {\bf \Gamma}^{\dag\sigma_1}_{o} 
{\bf \Lambda}^{2\alpha}_{e} 
{\bf \Gamma}^{\sigma_1}_{o} {\bf \Lambda}_{o}
\right)
{\bf \Gamma}^{\sigma_2}_{e} {\bf \Lambda}_{e}
\right)
\cdots
{\bf \Gamma}^{\sigma_{\ell}}_{e/o} {\bf \Lambda}_{e/o}
\right],
\ee
where we are using as computational basis the eigenvectors of the local operators $s^{z}_{j}$; the total computational cost being $O(\ell\chi^3)$.

\begin{figure}
    \centering
\includegraphics[width=\linewidth]{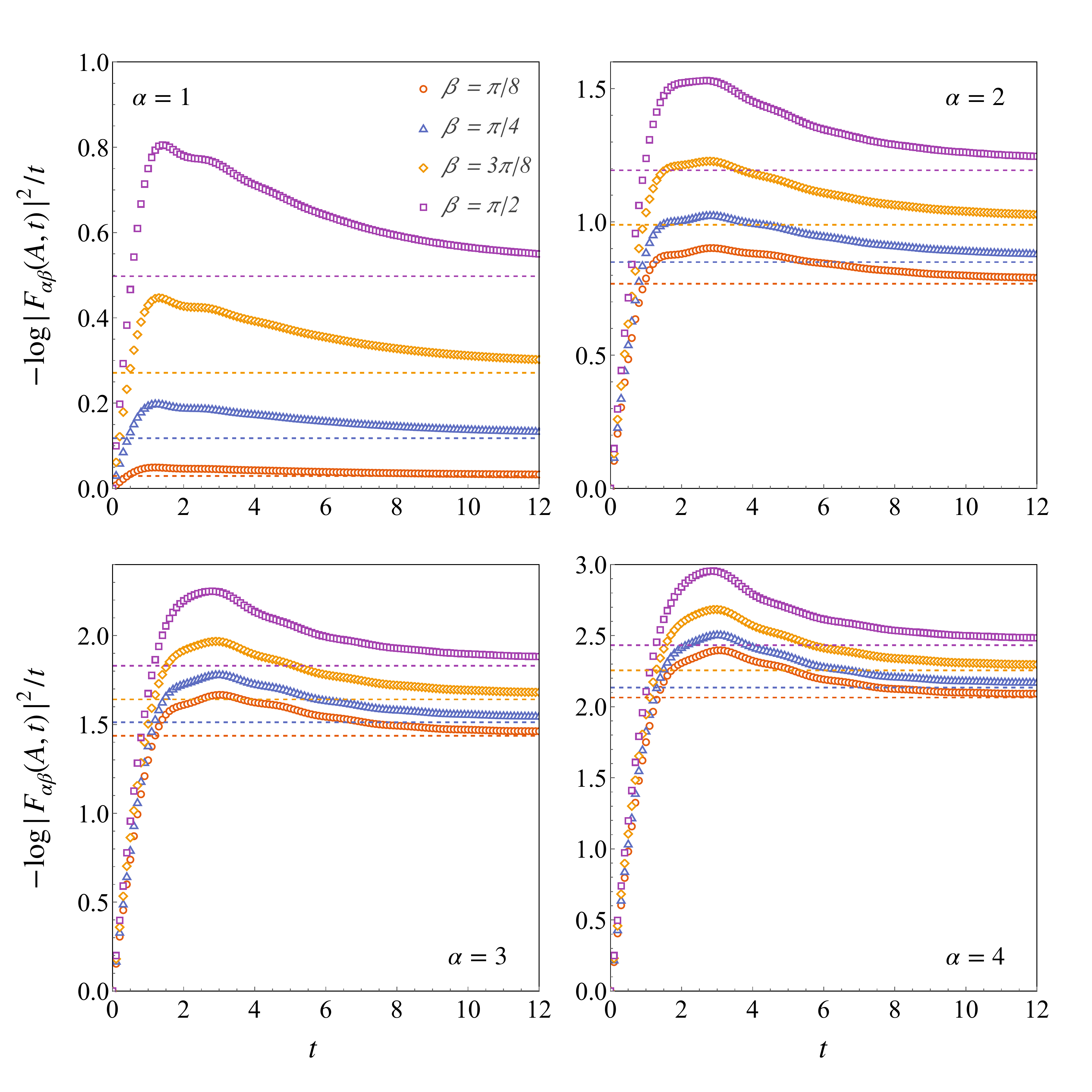}
    \caption{Logarithmic slope of the moment generating function $F_{\alpha,\beta}(A,t)$ after a quench in the XXZ model with $\Delta=1/2\, (p=2)$, starting from the N\'eel state. Symbols are the iTEBD data computed with $|A|=50$, dashed lines are the asymptotic predictions. 
    Different $\alpha$ values correspond to different panels; different symbols identify different values of $\beta$.}
    \label{fig:neel_p2}
\end{figure}

\begin{figure}
    \centering
\includegraphics[width=\linewidth]{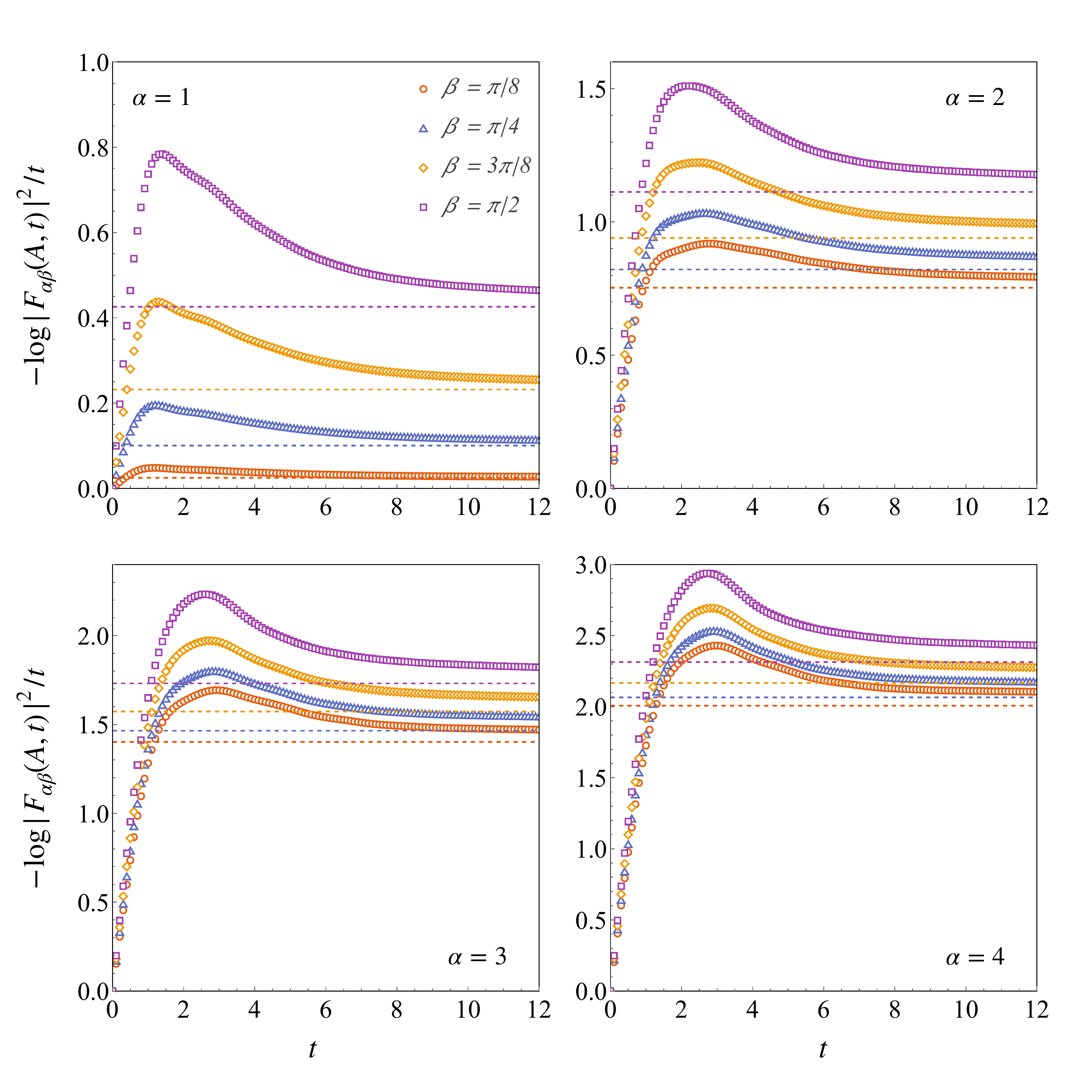}
    \caption{Same as in Figure~\ref{fig:neel_p2} for a quench in the XXZ model with $\Delta=1/\sqrt{2}\, (p=3)$, starting from the N\'eel state.}
    \label{fig:neel_p3}
\end{figure}

\begin{figure}
    \centering
\includegraphics[width=\linewidth]{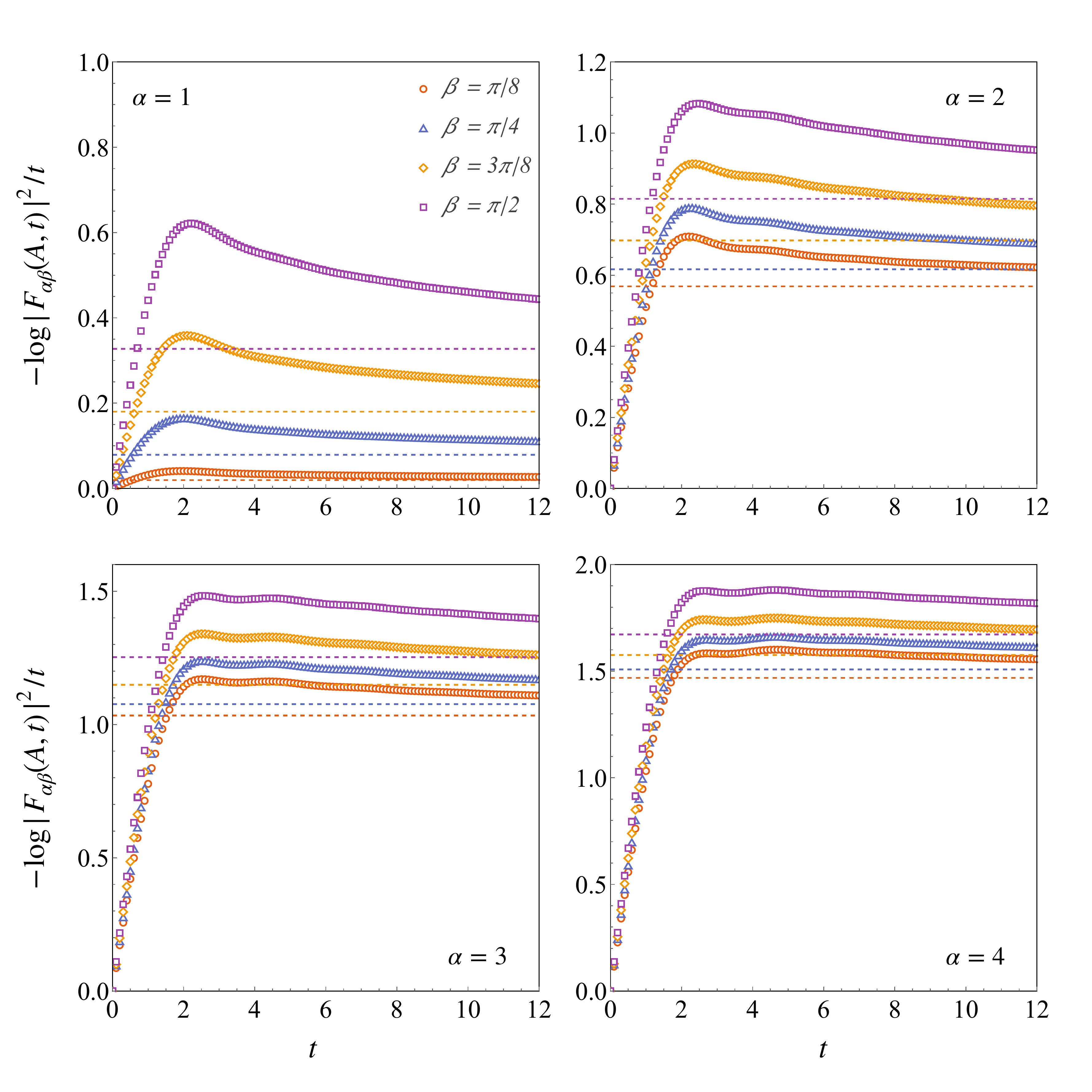}
    \caption{Logarithmic slope of the moment generating function $F_{\alpha,\beta}(A,t)$ after a quench in the XXZ model with $\Delta=1/2\, (p=2)$, starting from the Majumdar-Gosh state. Symbols are the iTEBD data computed with $|A|=50$, dashed lines are the asymptotic predictions. 
    Different $\alpha$ values correspond to different panels; different symbols identify different values of $\beta$.}
    \label{fig:mg_p2}
\end{figure}

\begin{figure}
    \centering
\includegraphics[width=\linewidth]{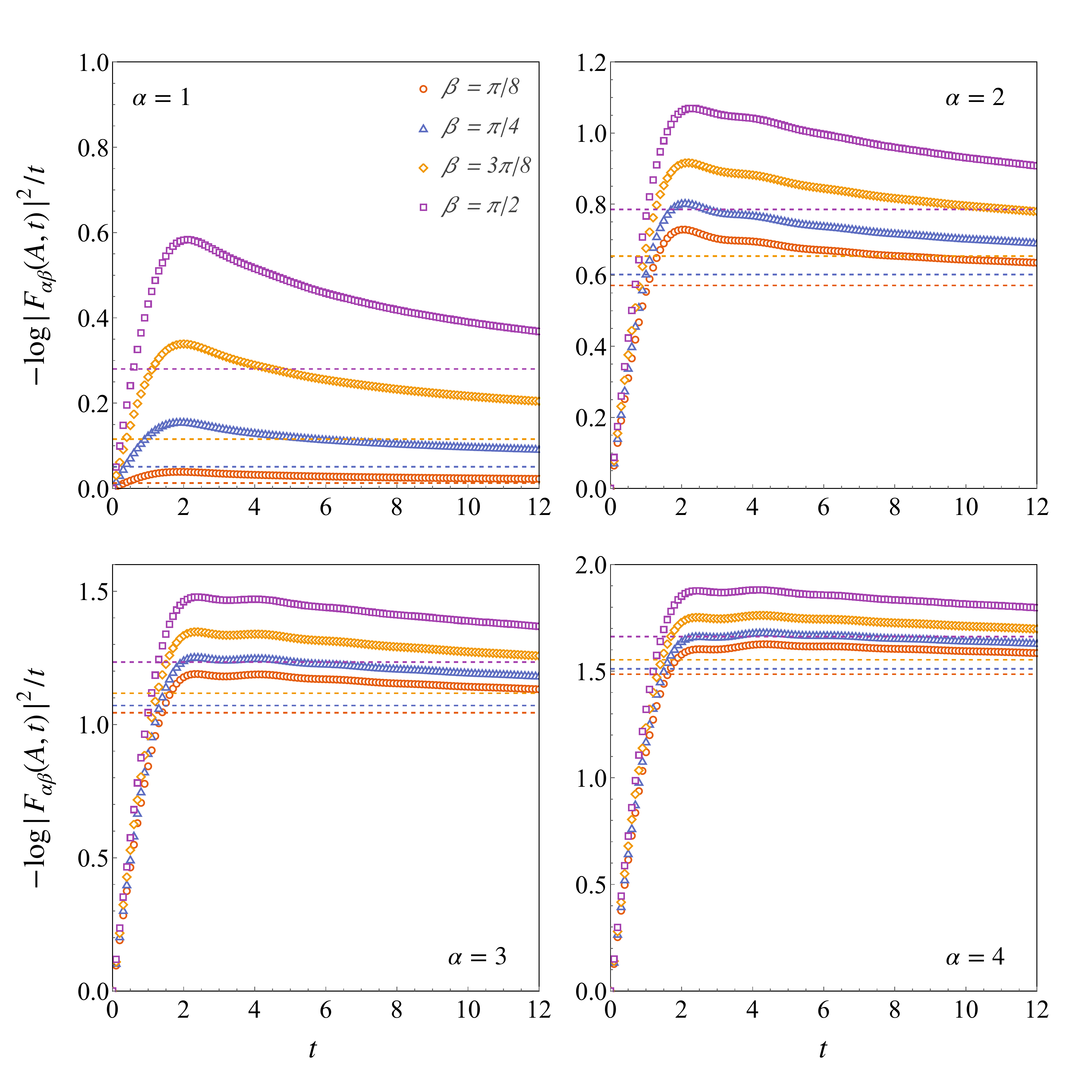}
    \caption{Same as in Figure~\ref{fig:mg_p2} for a quench in the XXZ model with $\Delta=1/\sqrt{2}\, (p=3)$, starting from the Majumdar-Gosh state.}
    \label{fig:mg_p3}
\end{figure}

It is easy to show, following similar arguments which lead to Eq.~(\ref{eq:chargemomentprediction}), that $F_{\alpha,\beta}(A,t) = e^{i \beta \bar q A} {\rm tr}[\tilde \rho^\alpha_{{\rm st},t} e^{i \beta \tilde Q_t}]{\rm tr}[\tilde \rho_{{\rm st},t} e^{- i \beta \tilde Q_t^T}]$, when $|A| > 2t$. 
Both $F_{\alpha,\beta}(A,t)$ and $Z_{\alpha,\beta}(A,t)$ decay exponentially toward their stationary values, with logarithmic slopes which are related via
$\log\left|F_{\alpha,\beta}(A,t)\right|^2 / t = \log \left[Z_{\alpha,\beta}(A,t)Z^{*}_{\beta}(A,t)\right] / t$. 

Further comparison between asymptotic predictions and iTEBD data are reported in Figures~\ref{fig:neel_p2} and \ref{fig:neel_p3} for quenches from the N\'eel state.
In this case, for sake of clarity, we report the slope of the generating functions of the local magnetisation computed in the semi-infinite chain, namely $F_{\alpha,\beta}(A,t)$. Indeed, we noticed that this quantity has a stronger dependence on $\beta$, which turns into more separated curves when plotting it for different values of $\beta$.

The agreement between exact numerical simulations and asymptotic prediction is surprisingly good for $p=2$ and all values of parameters we are considering. On the contrary, when quenching toward $\Delta=1/\sqrt{2}\, (p=3)$ the approaching to the asymptotic predictions seem much slower.

Similar considerations can be done for the quenches from the Majumdar-Gosh state, whose comparison with the asymptotics are reported in Figures~\ref{fig:mg_p2} and \ref{fig:mg_p3} for $p=2$ and $p=3$ respectively.

\end{document}